\documentclass[resetfootnotes,trackchanges]{aastex701}
\received{2026 July 3}
\revised{2026 September 7}
\accepted{2026 October 2}

\usepackage{amsmath}
\usepackage{gensymb}
\usepackage{verbatim}
\usepackage{siunitx}

\begin{document}

\title{First L-type Substellar Object Discovered in the Vicinity of the Embedded Cluster NGC\,2023}

\author[orcid=0000-0002-5378-2240,sname='Styliani Tsilia']{Styliani Tsilia}
\affiliation{Instituto de Astrofísica de Canarias, Vía Láctea, 38205, La Laguna, Tenerife, Spain}
\affiliation{Universidad de La Laguna, Departamento de Astrofísica, 38206, La Laguna, Tenerife, Spain}
\email[show]{stsilia@iac.es}  

\author[orcid=0000-0002-0463-6212,sname='Carlos Dominguez-Tagle']{Carlos Dominguez-Tagle}
\affiliation{Instituto de Astrofísica de Canarias, Vía Láctea, 38205, La Laguna, Tenerife, Spain}
\affiliation{Universidad de La Laguna, Departamento de Astrofísica, 38206, La Laguna, Tenerife, Spain}
\email{} 

\author[orcid=0000-0001-6023-4974,sname='Maruša Žerjal']{Maruša Žerjal}
\affiliation{Instituto de Astrofísica de Canarias, Vía Láctea, 38205, La Laguna, Tenerife, Spain}
\affiliation{Universidad de La Laguna, Departamento de Astrofísica, 38206, La Laguna, Tenerife, Spain}
\email{}

\author[orcid=0000-0002-1208-4833,sname='Eduardo L. Martín']{Eduardo L. Martín}
\affiliation{Instituto de Astrofísica de Canarias, Vía Láctea, 38205, La Laguna, Tenerife, Spain}
\affiliation{Universidad de La Laguna, Departamento de Astrofísica, 38206, La Laguna, Tenerife, Spain}
\email{}  

\author[orcid=0000-0002-7084-487X,sname='Hervé Bouy']{Hervé Bouy}
\affiliation{Laboratoire d’Astrophysique de Bordeaux, Univ. Bordeaux, CNRS, B18N, allée Geoffroy Saint-Hilaire, 33615 Pessac, France}
\affiliation{Institut universitaire de France (IUF), 1 rue Descartes, 75231, PARIS CEDEX 05, France}
\email{} 

\author[orcid=0009-0001-9574-8585,sname='Nafise	Sedighi']{Nafise	Sedighi}
\affiliation{Instituto de Astrofísica de Canarias, Vía Láctea, 38205, La Laguna, Tenerife, Spain}
\affiliation{Universidad de La Laguna, Departamento de Astrofísica, 38206, La Laguna, Tenerife, Spain}
\email{} 

\author[orcid=0000-0003-4269-4779,sname='Sara Muñoz Torres']{Sara Muñoz Torres}
\affiliation{Instituto de Astrofísica de Canarias, Vía Láctea, 38205, La Laguna, Tenerife, Spain}
\affiliation{Universidad de La Laguna, Departamento de Astrofísica, 38206, La Laguna, Tenerife, Spain}
\email{} 

\author[orcid=0000-0003-2150-0787,sname='Nikola Vitas']{Nikola Vitas}
\affiliation{Instituto de Astrofísica de Canarias, Vía Láctea, 38205, La Laguna, Tenerife, Spain}
\affiliation{Universidad de La Laguna, Departamento de Astrofísica, 38206, La Laguna, Tenerife, Spain}
\email{} 

\author[orcid=0000-0003-1464-9276,sname='Khalid	Barkaoui']{Khalid	Barkaoui}
\affiliation{Instituto de Astrofísica de Canarias, Vía Láctea, 38205, La Laguna, Tenerife, Spain}
\affiliation{Universidad de La Laguna, Departamento de Astrofísica, 38206, La Laguna, Tenerife, Spain}
\affil{Astrobiology Research Unit, Université de Liège, 19C Allée du 6 Août, 4000 Liège, Belgium}
\affil{Department of Earth, Atmospheric and Planetary Science, Massachusetts Institute of Technology, 77 Massachusetts Avenue, Cambridge, MA 02139, USA}
\email{} 

\author[orcid=0000-0002-3612-8968,sname='Nicolas Lodieu']{Nicolas Lodieu}
\affiliation{Instituto de Astrofísica de Canarias, Vía Láctea, 38205, La Laguna, Tenerife, Spain}
\affil{Universidad de La Laguna, Departamento de Astrofísica, 38206, La Laguna, Tenerife, Spain}
\email{}

\author[orcid=0000-0001-5392-2701,sname='Jerry J.-Y. Zhang']{Jerry J.-Y. Zhang}
\affiliation{Department of Physics and Astronomy, The University of Western Ontario, 1151 Richmond Street, London, ON N6A 3K7, Canada}
\affil{Institute for Earth and Space Exploration, The University of Western Ontario, 1151 Richmond Street, London, ON N6A 3K7, Canada}
\email{}

\begin{abstract}

We present the first \textcolor{black}{substellar object in a star-forming region spectroscopically confirmed with Euclid, namely J054158.70-021031.4.} This young and faint source \textcolor{black}{is located close to the embedded open cluster NGC\,2023. It has a} spectral type \textcolor{black}{of} L3, \textcolor{black}{a} reddening \textcolor{black}{of} $A_\mathrm{{V}}=3$\,mag, \textcolor{black}{proper motion consistent with the nearby cluster NGC\,2024,} and a magnitude of ${\sim}24.5$ in the Euclid optical band. \textcolor{black}{At the distance of 403\,pc, its inferred mass lies close to the planetary-mass boundary.} We discovered the object by searching for ultracool dwarf candidates in the Euclid Early Release Observations. The search was based on photometric criteria, followed by the analysis of slitless 2D spectroimages in order to distinguish features that indicate the nature of these objects. This work demonstrates the potential of Euclid to unveil previously hidden populations of very low-mass objects, located in obscured regions where \textcolor{black}{star formation is ongoing.}   

\end{abstract}



\section{Introduction}
Beyond the hydrogen-burning mass limit set at ${\sim} 0.08\,M_\odot$ \citep{1963PThPh..30..460H, 1963ApJ...137.1121K}, toward the very low-mass end of the mass spectrum, we encounter the regime of substellar objects. These sources include brown dwarfs and free-floating planetary-mass objects. Together with very low-mass stars, they comprise the category of sources known as ultracool dwarfs (UCDs), which are objects of spectral type M7 and later \citep{1997AJ....113.1421K} corresponding to effective temperatures below ${\sim}2700$\,K \citep{2015ApJ...810..158F}. UCDs are difficult to observe due to their intrinsic faintness, and many open questions persist about their exact formation mechanisms, their evolution, and atmospheric properties.

Yet these objects are highly significant; acting as a bridge between stars and planets, they are numerous both in young open clusters and the field \citep{penaramirezNEWISOLATEDPLANETARYMASS2012, 2021ApJS..253....7K} and therefore essential to construct a complete picture of star formation processes. Their cool temperatures allow molecular species and grains of dust to survive in their atmospheres \citep{2005ARA&A..43..195K}. Moreover, understanding how the initial mass function (IMF) extends into the very low-mass regime \cite[e.g.,][]{2024ApJS..271...55K}, and how it might be influenced by environmental factors \cite[e.g.,][]{2025MNRAS.537..752B}, is crucial for connecting the core mass function (CMF) to the final outcome of star formation. 

Young star-forming regions provide a crucial laboratory for probing the very low-mass regime. UCDs keep fading over time, so young regions allow us to detect them at their brightest phase, reaching down to just a few Jupiter masses \cite[e.g.,][]{penaramirezNEWISOLATEDPLANETARYMASS2012}. Unlike field populations, young regions offer access to samples of objects before \textcolor{black}{dynamical evolution significantly modifies their spatial distribution.} Hence, these objects preserve information about their formation and can provide good constraints on the IMF. \textcolor{black}{Orion has been the target of numerous searches for very low-mass objects. Early work identified L-type members in $\sigma$\,Orionis already more than two decades ago \cite[see, e.g.,][]{2001ApJ...558L.117M}. Subsequent surveys extended the substellar census across several Orion populations, including Orion OB1a and OB1b \citep{2008AJ....136...51D}, the Trapezium Cluster \citep{2005MNRAS.361..211L, 2006MNRAS.373L..60L}, NGC\,2024 \citep{2006ApJ...646.1215L}, $\sigma$\,Orionis \citep{penaramirezNEWISOLATEDPLANETARYMASS2012}, and Barnard\,30 \citep{2026A&A...714A..30B}. In contrast, the local environment around NGC\,2023 and the Horsehead Nebula remains underexplored at substellar masses.} This relatively close (${\sim}400$\,pc) field is \textcolor{black}{a star-forming region} featuring a structure shaped by strong radiation fields and dense molecular material \citep{2005A&A...437..177H}. The region lies at the interface between molecular gas and the radiation from nearby massive stars, making it a prime example of an environment influenced by stellar feedback processes such as photoionization and radiation-driven compression. \textcolor{black}{The study of such regions can highlight the role of local cloud structure in shaping the lowest-mass products of star formation.}

One powerful tool in the search for very low-mass objects is the Euclid space mission. It has already unveiled thousands of UCD candidates and hundreds of new spectroscopically confirmed UCDs \citep{2025A&A...696A..80B, 2025ApJ...991...84D, martinEuclidEarlyRelease2025, 2026A&A...713A.245M, 2026A&A...711A..37Z}, demonstrating its sensitivity down to ${\sim}4$\,$M_\mathrm{Jup}$ at the distance of Orion \citep{martinEuclidEarlyRelease2025}, thanks to the performance of its Visible Camera \cite[VIS;\!\!][]{2025A&A...697A...2E} and Near-Infrared Spectrometer and Photometer \cite[NISP;\!\!][]{2025A&A...697A...3E}. Euclid's VIS instrument provides high-resolution imaging in a single broad optical band ($0.55–0.92\,\mu$m, hereafter referred to as $I_{\mathrm{E}}$) with an angular resolution of ${\sim}$0\farcs1. The second instrument, NISP, operates in the near-infrared, providing images in the $Y_{\mathrm{E}}$ ($0.95–1.21\,\mu$m), $J_{\mathrm{E}}$ ($1.17–1.57\,\mu$m), and $H_{\mathrm{E}}$ ($1.52–2.02\,\mu$m) photometric bands, and slitless spectroscopy over $0.93–1.89\,\mu$m with a resolving power $R\approx450$. Euclid's Wide Survey reaches nominal limiting magnitudes (AB, $5\sigma$ point-source depth) of 24.5\,mag in the NISP bands and 26.2\,mag in the VIS band (\!\!\citealt{2022A&A...662A.112E}). Euclid's Deep Survey is expected to reach 2 magnitudes deeper (\!\!\citealt{2021A&A...655A..44E}). Euclid delivers a wide field of view (FoV) per pointing, with footprints of ${\sim}0.54$\,deg$^2$ for VIS (\!\!\citealt{2025A&A...697A...2E}) and ${\sim}0.57$\,deg$^2$ for NISP (\!\!\citealt{2025A&A...697A...3E}).

In this work, we exploit for the first time Euclid observations to successfully identify an L3-type substellar object in an embedded region, where both reddening and negligible proper motions limit conventional searches. We take the first step toward probing the substellar population of this challenging environment, pushing the low-mass limit of the known population toward the planetary-mass regime.

\section{Data} 
 
This paper utilizes the Euclid Early Release Observations (ERO) program. More specifically, it focuses on the Barnard\,33 (Horsehead Nebula) field, one of the pointings included in project ERO-02\footnote{https://euclid.esac.esa.int/dr/ero/} (P.I.: Eduardo Martín), which targeted nearby star-forming regions (distance $\leq400$\,pc) and very young open clusters (age $\leq10$\,Myr).
 
In this work, we used photometric data from Euclid, processed by the pipeline of \citet{2025A&A...697A...6C}, as well as a denebulized photometric catalog. The denebulized catalog was constructed by applying a novel deep learning algorithm (D\textsc{e}N\textsc{eb}; E. Bertin et al., in preparation) able to distinguish between the nebula and point source components in the images 
\cite[for a more detailed description of this algorithm, see][]{2026A&A...706A.149D}. We complemented Euclid data with ancillary observations from the Spitzer mission \citep{2004ApJS..154....1W} in the IRAC \textcolor{black}{$[3.6]$ and $[4.5]$ channels with central wavelengths of $3.56$ and $4.52\,\mu$m, respectively \citep{2004ApJS..154...10F}}, which we processed with D\textsc{e}N\textsc{eb} as well. Denebulized data is explicitly noted in the plots by using D\textsc{e}N\textsc{eb}'s name next to the name of each denebulized band.

\section{Candidate selection and confirmation} \label{sec:three}
Our search focused only on the high-reddening region of the Euclid ERO Barnard\,33 field; the low-reddening area has been studied by \citet{martinEuclidEarlyRelease2025} and C. Dominguez-Tagle et al. (in preparation). We set a constant R.A. value of 85.18750\degree{} as the boundary between these two regions. We filtered the denebulized photometric catalog, requiring a signal-to-noise ratio (S/N)\,$\geq3$ in the $J_{\mathrm{E}}$ band (where we expect substellar objects to be the brightest). This selection cut was kept minimal and did not include morphological criteria, to ensure we would not accidentally filter out true members of the star-forming region. Subsequently, we constructed a grid of color--magnitude diagrams (CMDs) combining data from Euclid with Spitzer, including only objects with valid photometry in each individual plot, and overlaying ATMO isochrones \citep{2020A&A...637A..38P} for 1--7\,Myr, consistent with the range of ages reported in the literature for this region \cite[see][and references therein]{2013MNRAS.429..775L}. The models were converted from the Vega to the AB photometric system and calculated for a distance of \textcolor{black}{$403\pm5$\,pc.} This distance was based on the value that \citet{2026A&A...708A.299D} calculated for NGC\,2024 and was considered as the most reliable proxy since NGC\,2023 and NGC\,2024 are part of the same cluster grouping within the Orion B molecular cloud. As reddening might shift the colors of sources toward the red part of the CMDs, coeval objects did not necessarily lie along the same isochrone. Therefore, in order to compile a list of preliminary candidates, we counted every object consistently redder than an isochrone in every CMD where the object appeared. We repeated this for every isochrone, also retaining sources lying up to 0.25\,mag blueward of each isochrone to account for uncertainties. This margin was based on the CMD position of the benchmark objects listed in \citet{martinEuclidEarlyRelease2025}. \textcolor{black}{This process yielded a preliminary list of 3148 candidates.}

Given the particularities of this field (i.e., nebulae, reddening, and artifacts), we filtered out the list of candidates to accelerate the selection by reviewing the NISP slitless 2D spectroimages. In the wavelength range of NISP, the spectra of substellar objects show characteristic strong absorption bands of H$_2$O. This feature translates into a drop in brightness that is evident even in the 2D spectroimages. The list of candidates for spectra extraction was limited to ${\sim}20.5$\,mag in the $J_{\mathrm{E}}$ band, in order to obtain spectra where the depth of H$_2$O absorption bands had a S/N\,$>$\,10, as described in C. Dominguez-Tagle et al. (in preparation). We produced image cutouts for each object and carried out a visual inspection of the spectral traces. We selected those that clearly exhibited substellar-like features in order to be extracted. The details of this method of candidate selection and prioritization through visual examination are included in S. Tsilia et al. (in preparation), which describes the full analysis of all the objects in this field. Even though this visual technique could include some bias, it has allowed us to go through the  artifacts and contamination present in slitless spectroscopy. It has also been very useful in a highly populated region with the presence of nebulae. It has allowed us to confirm or reject the UCD nature of candidates, given the available data in the most efficient way possible.

\begin{figure*}
\centering
\includegraphics[width=0.505\linewidth]{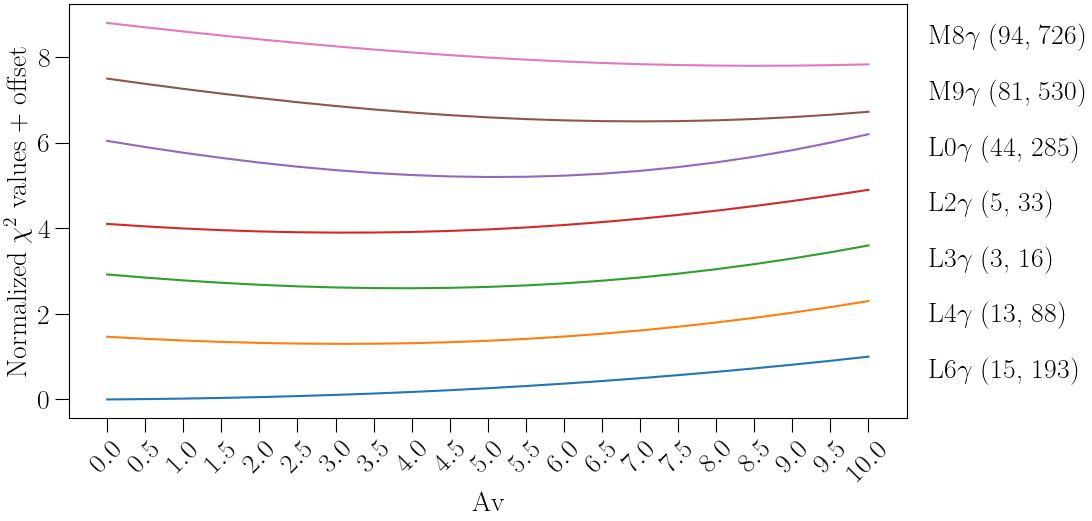}
\includegraphics[width=0.45\linewidth]{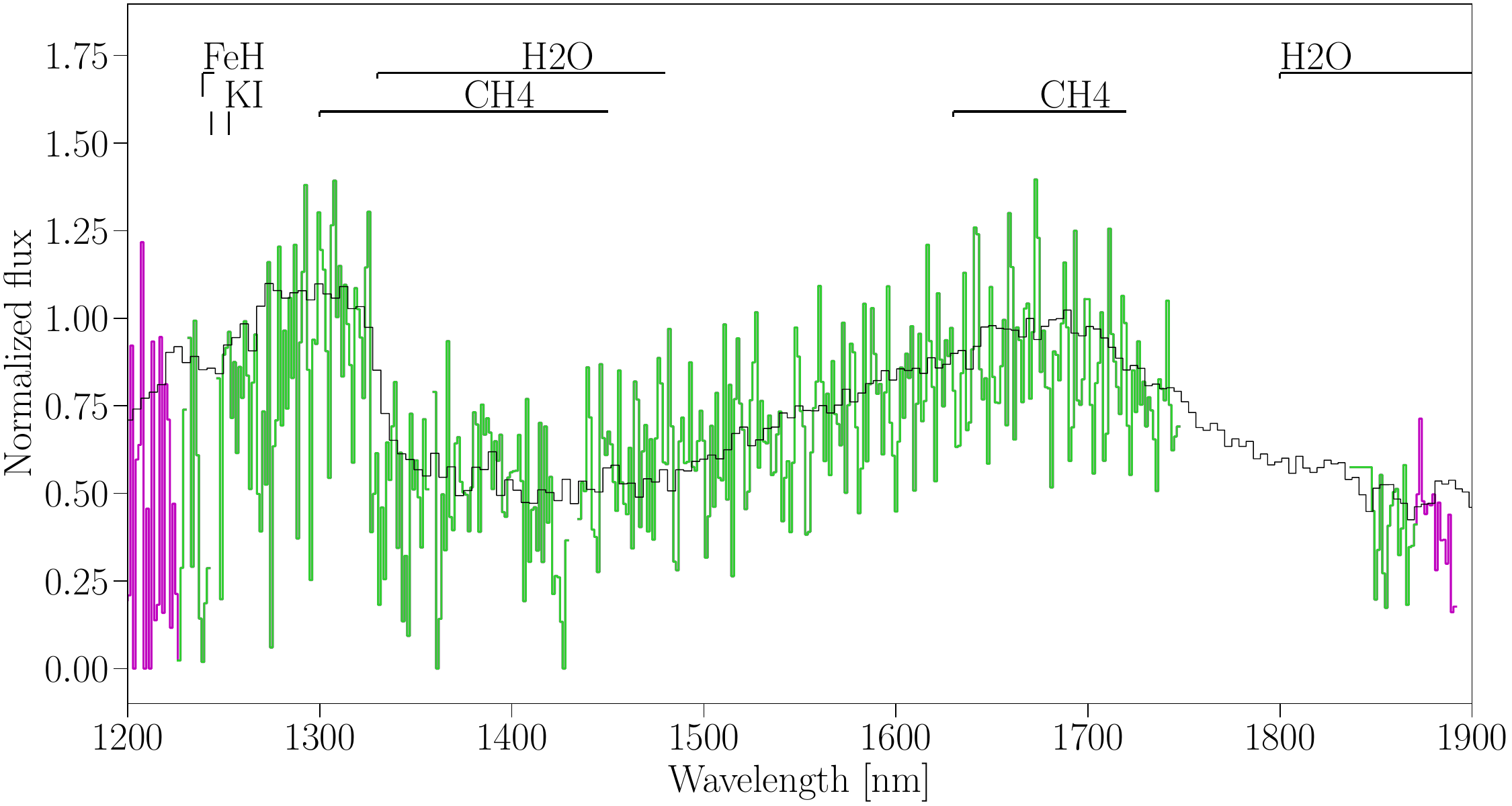}
\caption{Left:  \textcolor{black}{$\chi^2$ values for the comparison of the smoothed J0541 Euclid spectrum with young templates at different $A_\mathrm{{V}}$ values. The $\chi^2$ values are normalized between each template minimum and maximum, which are written next to the template spectral type.} Right: \textcolor{black}{the Euclid spectrum of J0541} dereddened for $A_\mathrm{{V}}=3$\,mag compared to an L3$\gamma$ template. The spectrum is plotted by green and magenta lines, the latter corresponding to wavelengths of low grism transmission. No line is plotted where the spectrum is contaminated by artifacts. The L3$\gamma$ template \textcolor{black}{ \cite[2MASSW J2208136+292121; ][]{gagne15} is plotted by the black line. The J0541 Euclid spectrum is available as the data behind the figure.
(The data used to create this figure are available in the
\href{https://doi.org/10.3847/1538-4357/aeb0a2}{online article}.)}
}
\label{fig:1dspe}
\end{figure*}

In the process of reviewing the candidates, the first we found to showcase the characteristic UCD spectral features was at coordinates \textcolor{black}{J054158.70-021031.4.} \textcolor{black}{Hereafter, we refer to the source as J0541 for brevity.} We then extracted the 1D spectrum. As described in \!\!\cite{2022A&A...662A.112E}, Euclid observations include four spectra by performing a dithering pattern. However, given that NISP has a mosaic of 16 detectors with gaps between them, all or part of the object spectrum could fall into these gaps for some of these dithering positions. In consequence, for a given object, the available spectra could go from zero to four. In the case of \textcolor{black}{J0541}, only one dithering position was available. We extracted the 1D spectrum using the collection of programs described in C. Dominguez-Tagle et al. (in preparation), which included the trace identification, the elimination of bad pixels, the subtraction of background, and a box-car extraction aperture. We did the wavelength calibration using the cut-on and cut-off wavelengths of the red grism and fitting to  the spectral dispersion measured by \!\!\citet{Gillard2026}. We corrected the instrumental response  by the sensitivity curve of the spectrophotometric standard star GRW70 observed in the same NISP configuration as that of \textcolor{black}{J0541}. These first-order calibrations were accurate enough to proceed with the classification by spectral type (SpT). We dereddened the spectrum for a range of $A_\mathrm{{V}}$ values from 0 to 10\,mag with a step of 0.5\,mag, using the \citet{1989ApJ...345..245C} extinction law. We compared all these dereddened spectra \textcolor{black}{(smoothed with a boxcar = 3)} to SpeX low-gravity standard templates \citep{SPEX, SPLAT} by $\chi^2$ minimization. The result is shown in the left panel of Figure\,\ref{fig:1dspe}. The \textcolor{black}{plot} shows the $\chi^2$ values for the comparison with all spectral types and extinction values. The minimum value is found for an $A_\mathrm{{V}}=3$\,mag and the L3$\gamma$ template 
\textcolor{black}{\cite[2MASSW J2208136+292121; ][]{gagne15}. For the other standard templates, the minimum, although higher than that for  L3$\gamma$, shifts to lower  $A_\mathrm{{V}}$ values for later types and to higher ones for earlier types.}
We did the same comparison with field standard templates from the same catalogs, obtaining 3 times larger values for $\chi^2$, which confirmed that the best fit corresponds to low-gravity standard templates.
\textcolor{black}{We computed the $H$-cont index defined by \citet{Allers13} to verify the triangular shape of the $H$-band peak, which is representative of young objects. For the original spectrum, the $H$-cont index is 1.091 $\pm$ 0.077, and for the spectrum dereddened for $A_\mathrm{{V}}=3$\,mag, the $H$-cont index is 1.080 $\pm$ 0.071, which confirms the triangular shape.}
The right panel of Figure\,\ref{fig:1dspe} shows the \textcolor{black}{J0541 spectrum dereddened for $A_\mathrm{{V}}=3$\,mag, } compared to the L3$\gamma$ template. Owing to contamination from artifacts, the flux for some wavelength values is masked. This spectrum confirms the candidate as a young L3 UCD. 

\section{The new object} 

The newly discovered substellar object exhibits very low brightness, being almost as faint as 24.5\,mag in the optical band, but becoming brighter as we move further into the infrared regime. A comparison of its photometry between the ERO pipeline and the denebulized catalog reveals that D\textsc{e}N\textsc{eb} primarily affected the $Y_{\mathrm{E}}$-band photometry, resulting in redder $Y_{\mathrm{E}}-J_{\mathrm{E}}$, and $Y_{\mathrm{E}}-H_{\mathrm{E}}$ colors by ${\sim}0.4$\,mag. This suggests that the deep learning algorithm successfully removed the contamination caused by nebulosity. The $I_{\mathrm{E}}$, $J_{\mathrm{E}}$, and $H_{\mathrm{E}}$ measurements remain essentially the same. This comparison is seen in Table\,\ref{tab:phot}, where additional photometric measurements for this object from \textcolor{black}{UKIDSS DR9 \citep{2007MNRAS.379.1599L}, the Spitzer Legacy catalog SESNA \citep{2019AAS...23336710G}, and VISTA/VIRCAM \cite[the Science Verification Galactic mini-survey of Orion:][as processed by \citealt{2026A&A...706A.210R}]{2010Msngr.139....6A, 2011Msngr.145...29P}} are combined.

Using VOSA \citep{2008A&A...492..277B}, we inspected the spectral energy distribution (SED) curve for \textcolor{black}{J0541} by combining the measurements exhibiting the lowest uncertainty: Euclid's denebulized photometry, data from VISTA\,Ks, and data from IRAC\,\textcolor{black}{$[3.6]$} and \textcolor{black}{$[4.5]$}. No obvious infrared excess was detected up to IRAC\,\textcolor{black}{$[4.5]$}. We also fitted ATMO 2020 \citep{2020A&A...637A..38P} chemical equilibrium models to the SED. For the gravity and effective temperature of the object, the best-fitting parameters were $\mathrm{log}g=2.50$, $T_\mathrm{{eff}}=1500\,\mathrm{K}$; the corresponding posterior 96\% credible limits were 2.50--2.95 and 1400--1770\,K, respectively. This temperature estimate is consistent, within its posterior 96\% credible interval, with the expected temperature of an L3-type, low-gravity object \citep{2015ApJ...810..158F}.

Additionally, we checked narrowband Br$\gamma$ observations that we had obtained with the PANIC wide-field near-infrared camera at the Calar Alto 2.2\,m telescope \citep{2018PASP..130b5003C} on 3 nights between 6 and 8 December 2025 (P.I.: Sara Muñoz Torres). \textcolor{black}{J0541} does not show obvious excess in the emission line. These observations will be further analyzed in future work.

\setlength{\tabcolsep}{16pt}
\begin{deluxetable*}{cccccc}
\tablewidth{0pt}
\tablecaption{Photometric measurements for \textcolor{black}{J0541}. \label{tab:phot}}
\tablehead{
\colhead{Band} & \colhead{D\textsc{e}N\textsc{eb}} & \colhead{ERO Pipeline} & \colhead{UKIDSS} & \colhead{VISTA} & \colhead{IRAC}
}
\startdata
$I_{\mathrm{E}}$ & $24.46\pm0.04$ & $24.48\pm0.05$ & \nodata & \nodata & \nodata \\
$Y_{\mathrm{E}}$ & $20.79\pm0.002$ & $20.38\pm0.02$ & \nodata & \nodata & \nodata \\
$J_{\mathrm{E}}$ & $19.48\pm0.001$ & $19.48\pm0.01$ & \nodata & \nodata & \nodata \\
$H_{\mathrm{E}}$ & $18.82\pm0.001$ & $18.77\pm0.01$ & \nodata & \nodata & \nodata \\
$Y$ & \nodata & \nodata & $20.06\pm0.21$ & \nodata & \nodata \\
$J$ & \nodata & \nodata & $18.58\pm0.10$ & $18.56\pm0.02$ & \nodata \\
$H$ & \nodata & \nodata & $17.40\pm0.06$ & $17.35\pm0.02$ & \nodata \\
$K$ & \nodata & \nodata & $16.47\pm0.05$ & \nodata & \nodata \\
$Ks$ & \nodata & \nodata & \nodata & $16.48\pm0.02$ & \nodata \\
\textcolor{black}{$[3.6]$} & \nodata & \nodata & \nodata & \nodata & $15.58\pm0.04$ \\
\textcolor{black}{$[4.5]$} & \nodata & \nodata & \nodata & \nodata & $15.43\pm0.05$ \\
\textcolor{black}{$[5.8]$} & \nodata & \nodata & \nodata & \nodata & $15.72\pm0.50$ \\
\enddata
\tablecomments{All values are in magnitudes. Euclid's bands are given in the AB system; all other bands in the Vega system. Whenever multiple data points are available for the same band, the measurement with the smallest uncertainty is displayed.
}
\end{deluxetable*}

\begin{figure*}[ht!]
\centering
\includegraphics[width=0.45\linewidth]{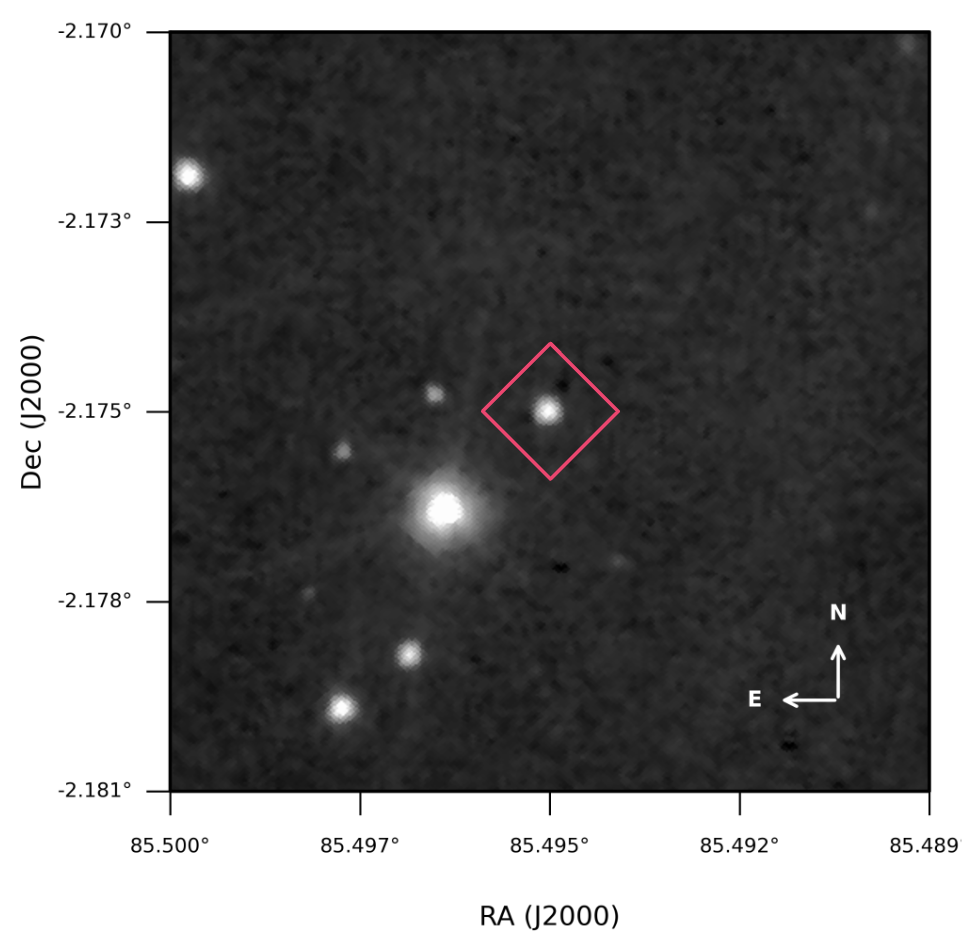}
\includegraphics[width=0.45\linewidth]{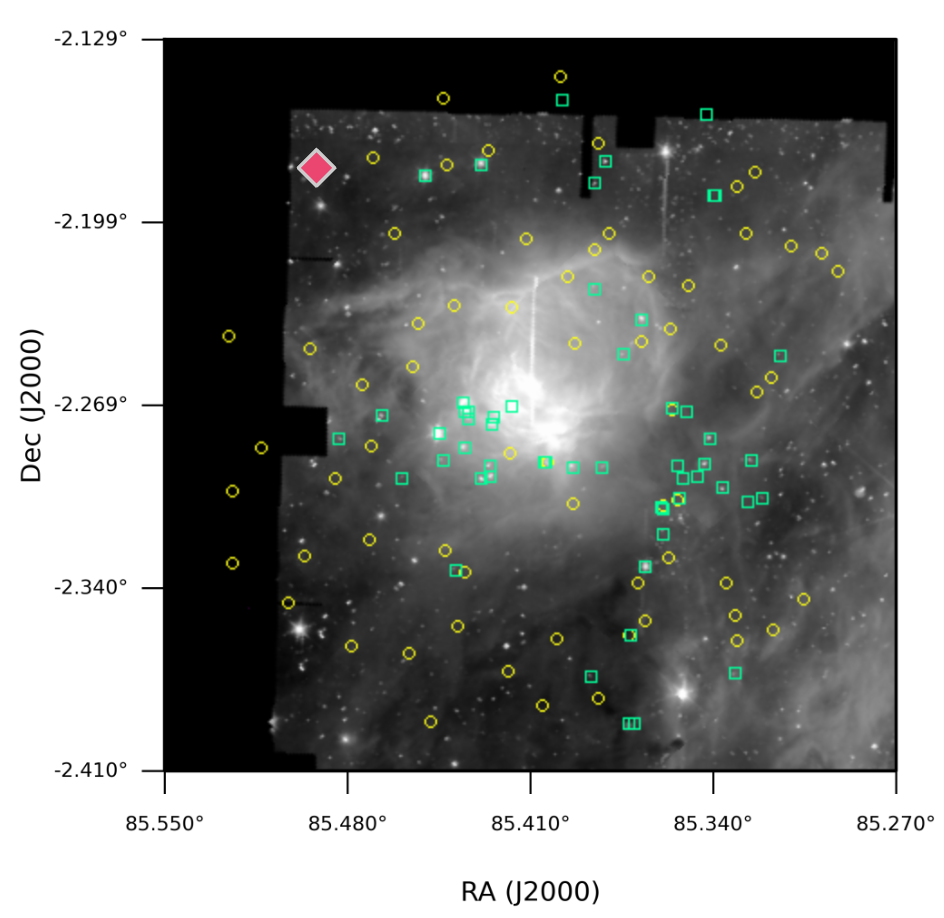}
\caption{Left: Euclid $J_{\mathrm{E}}$ cutout centered on \textcolor{black}{the source of interest (J054158.70-021031.4).} Right: spatial distribution of YSOs (denoted as green squares) and dense cores (denoted as yellow circles) falling within 8\arcmin{} of NGC\,2023. \textcolor{black}{J0541} is marked as a pink rhombus in both plots.
}
\label{fig:spatial}
\end{figure*}

\textcolor{black}{J0541} is located approximately 7\farcm5 away from the center of NGC\,2023, which translates to ${\sim}0.9$\,pc away from the cluster center at the distance of the cluster. An image cutout is given in the left panel of Figure\,\ref{fig:spatial}, where the object clearly appears as a point source. This region has been studied before, and several sources have been discovered in the same FoV. We note the presence of young stellar objects \cite[YSOs; see\!\!][]{2003yCat.2246....0C, 2009A&A...507.1485M, 2012AJ....144..192M, 2021A&A...649A...1G}, typical of a region undergoing star formation, and also the abundance of dense cores that seem almost evenly spread out \cite[see][]{2009A&A...507.1485M, 2016ApJ...817..167K, 2020ApJS..251...20D, 2020A&A...635A..34K}. Their spatial distribution is plotted in the right panel of Figure\,\ref{fig:spatial}. \citet{2009A&A...507.1485M} suggest that the expansion of the IC\,434 H\,II region (located just south of the Horsehead Nebula) compresses gas in dense, pillar-like structures, triggering star formation in this region.

\begin{figure*}[ht!]
\centering
\includegraphics[width=0.45\linewidth]{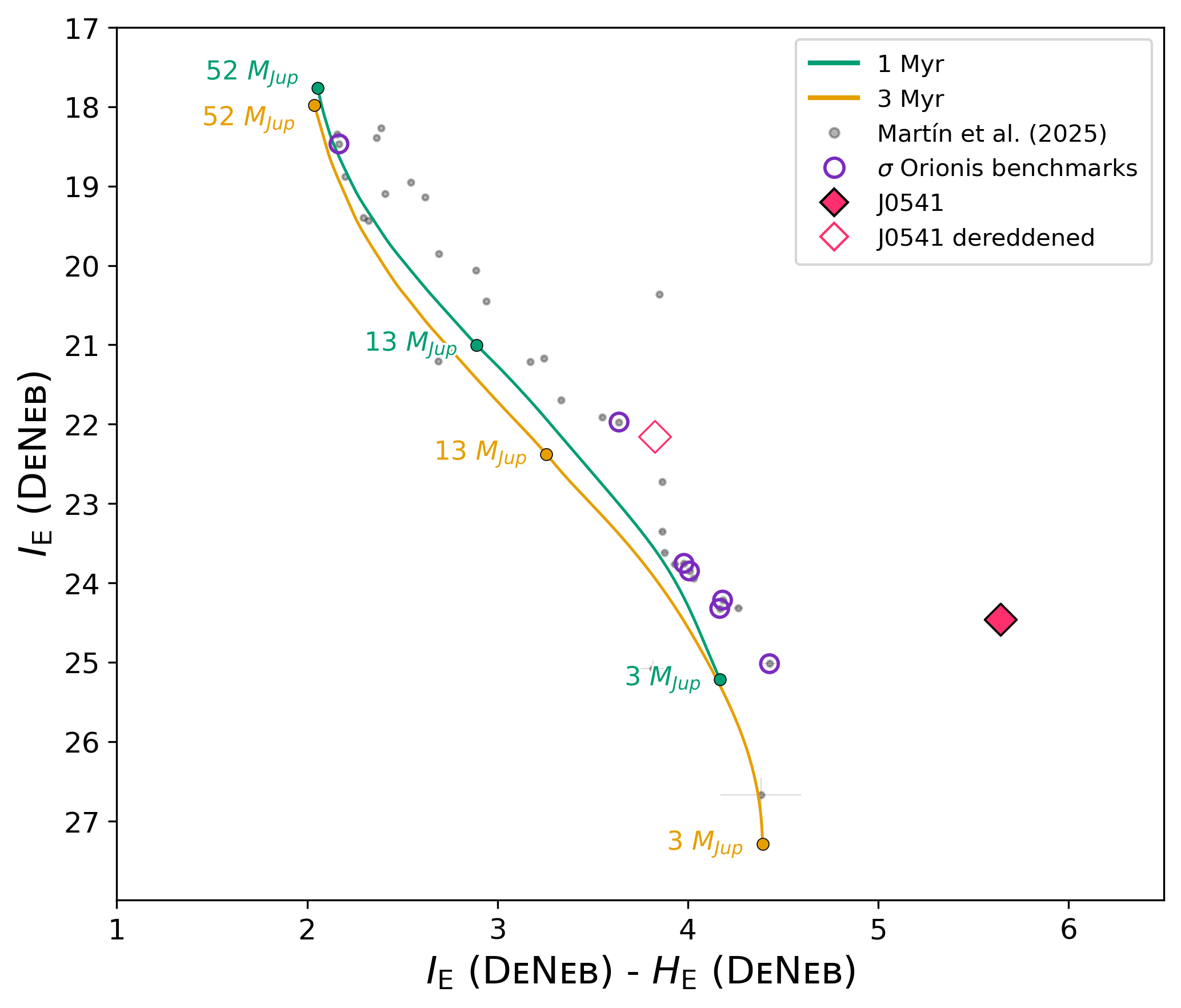}
\includegraphics[width=0.452\linewidth]{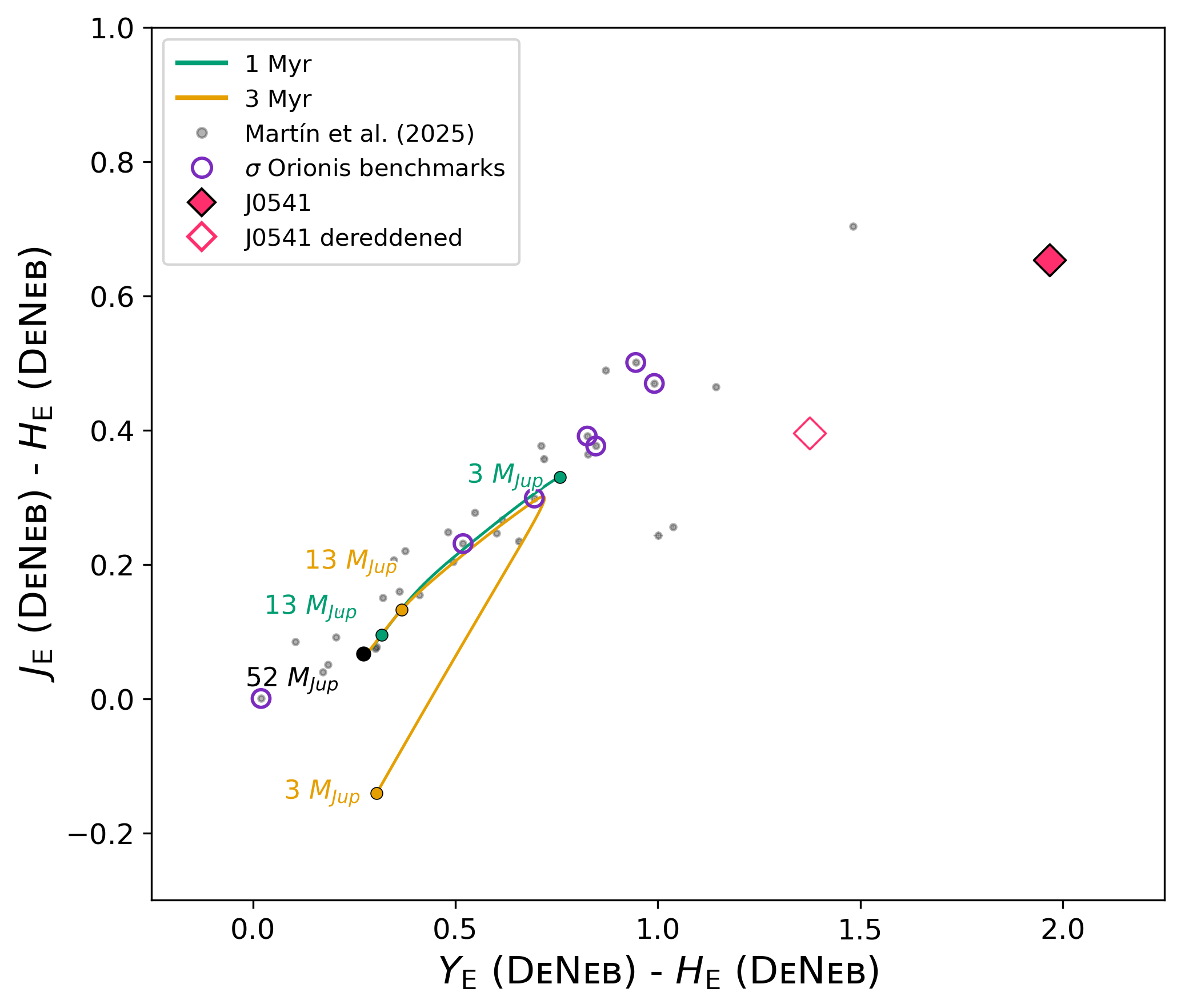}
\caption{Plots showing \textcolor{black}{J0541} compared to the new UCD candidates and previously known objects listed in \citet{martinEuclidEarlyRelease2025}. \textcolor{black}{J0541} is marked as a pink rhombus, benchmark objects are marked as purple circles, and the ATMO CEQ isochrones \cite[see][]{2020A&A...637A..38P} for the ages of 1 and 3\,Myr and a distance of 403\,pc are drawn. \textcolor{black}{J0541} is also shown dereddened using $A_\mathrm{{V}}=3$\,mag
and $R_\mathrm{{V}}=3.1$. Left: CMD of $I_{\mathrm{E}}$ versus $I_{\mathrm{E}}-H_{\mathrm{E}}$. Right: color--color diagram of $J_{\mathrm{E}}-H_{\mathrm{E}}$ versus $Y_{\mathrm{E}}-H_{\mathrm{E}}$. The 52\,$M_\mathrm{Jup}$ point overlaps for the two isochrones.}
\label{fig:cmd_cc}
\end{figure*}

Now we examine \textcolor{black}{J0541} relative to the new UCD candidates and previously known objects that \citet{martinEuclidEarlyRelease2025} reported in the low-reddening region of the Euclid ERO Barnard\,33 field. Additionally, we plot the same sources in color--color diagrams. An indicative CMD is shown in the left panel of Figure\,\ref{fig:cmd_cc}. \textcolor{black}{As mentioned in Section\,\ref{sec:three}, our full candidate selection considered isochrones spanning 1–7\,Myr, but here we show the 1 and 3\,Myr isochrones as representative models for the young population. The selection of J0541 is insensitive to the adopted age within the 1–7\,Myr range.} \textcolor{black}{J0541} is well separated to the right of the models, which is expected for a reddened object, and is almost as faint as the faintest sources detected in \citet{martinEuclidEarlyRelease2025}. We dereddened it using the value of $A_\mathrm{{V}}=3$\,mag that we found above and the \citet{1989ApJ...345..245C} extinction law. The dereddened source is in good agreement with the sequence of the rest of the objects. The plot also reveals that \textcolor{black}{J0541} is located close to the deuterium-burning mass limit of ${\sim}13$\,$M_\mathrm{Jup}$ \citep{2011ApJ...727...57S}, and could be of planetary mass, depending on its exact age. An indicative color--color diagram is shown in the right panel of Figure\,\ref{fig:cmd_cc}. Once again, we dereddened \textcolor{black}{J0541} using $A_\mathrm{{V}}=3$\,mag. We note that it lies further away from most of the other sources with respect to the isochrones, \textcolor{black}{but we caution that the theoretical isochrones do not accurately reproduce the observed colors and are shown only as a qualitative reference.}

\textcolor{black}{As an empirical assessment of the mass–extinction range accessible at the depth demonstrated by J0541, we used the 1 and 3\,Myr isochrones at 403\,pc to calculate the extinction required for objects of different masses to reach the observed $I_E=24.46$\,mag. Objects of ${\sim}$8, 13, 30, and 51\,$M_\mathrm{Jup}$ reach this reference magnitude at $A_V\simeq$\,0.9–3.0, 2.7–4.5, 6.8–7.3, and 8.4–8.7\,mag, respectively, while a 3\,Myr and ${\sim}$5\,$M_\mathrm{Jup}$ model is already fainter without extinction. Thus, at the depth of J0541, sensitivity to extinction increases strongly with mass and decreases with age at the lowest masses.}

\begin{figure*}[ht!]
\centering
\includegraphics[height=10cm]{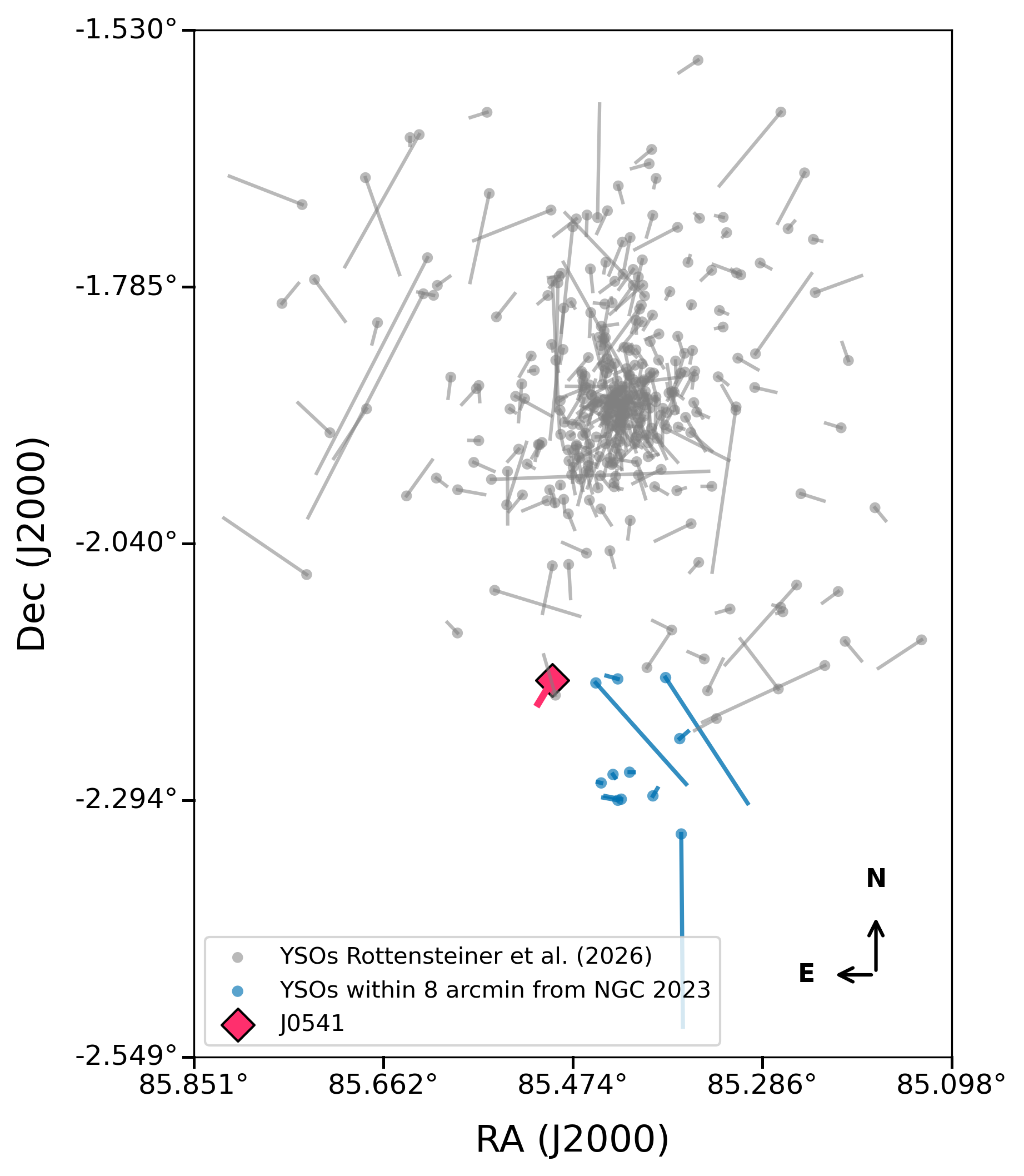}
\includegraphics[height=9.9cm]{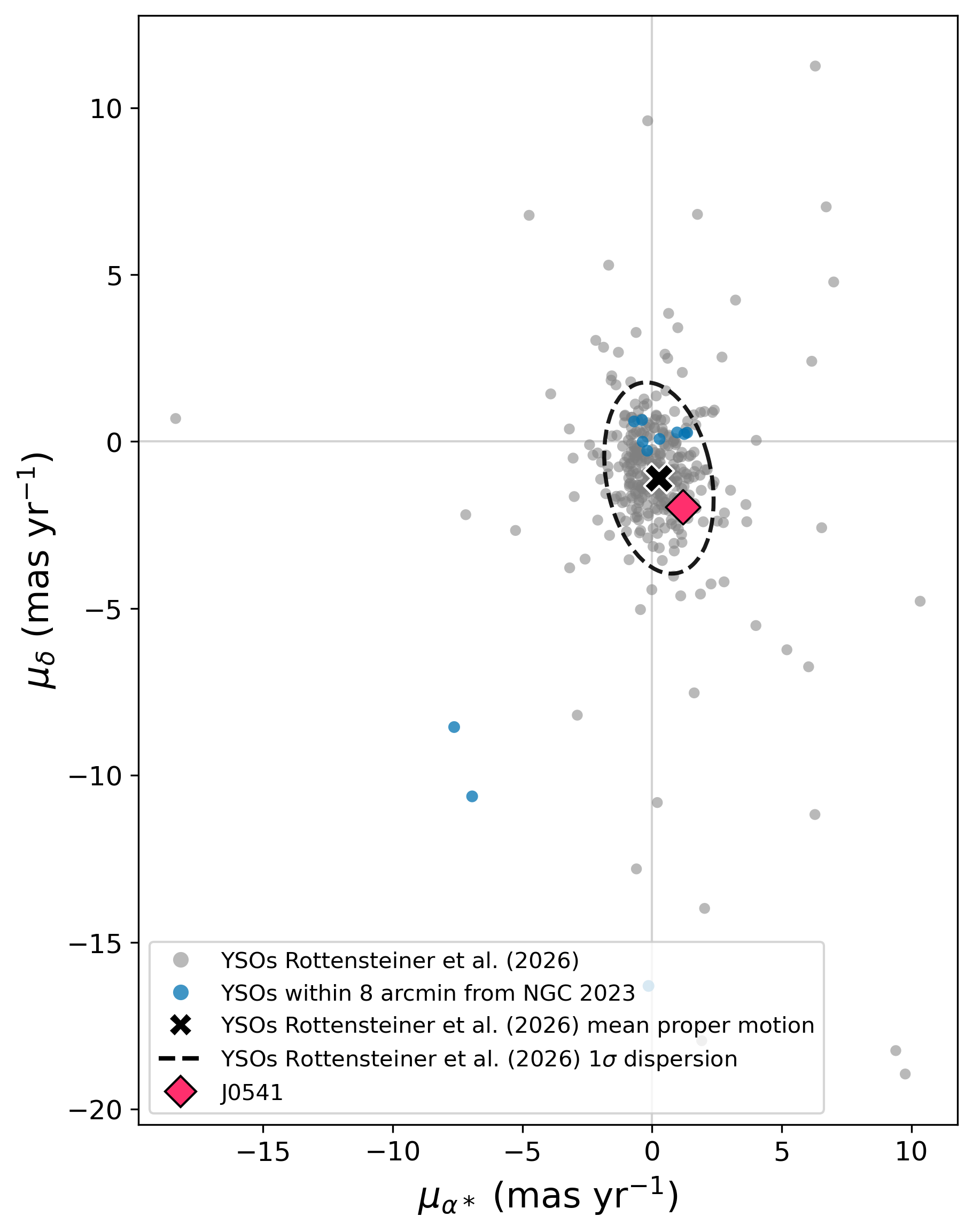}
\caption{Proper motion of \textcolor{black}{J0541} (marked as a pink rhombus) compared to proper motions of YSOs in the nearby environment. Members of NGC\,2024 are denoted in gray, while additional objects within 8\arcmin{} of the center of NGC\,2023 are denoted in blue. Only proper motions with S/N\,$\geq3$ are plotted. Left: the spatial distribution of sources is shown together with the direction of their proper motion. Velocities are shown in arbitrary units. Right: proper motion diagram. The black cross marks the mean proper motion of the cluster's members. The black dashed line denotes the 1$\sigma$ covariance ellipse. \textcolor{black}{The proper motion error for J0541 is smaller than the symbol.}}
\label{fig:motion}
\end{figure*}

The proper motion of this object was found by  \citet{2026A&A...706A.210R} to be ${\simeq}1.19\pm0.35\,\mathrm{mas\,yr^{-1}}$ in R.A. and ${\simeq}{-}1.97\pm0.21\,\mathrm{mas\,yr^{-1}}$ in decl., based on \textcolor{black}{multiepoch} infrared observations carried out with the VISTA/VIRCAM instrument \textcolor{black}{through three different surveys: the Science Verification Galactic mini-survey of Orion \citep{2010Msngr.139....6A, 2011Msngr.145...29P}, the Vista Hemisphere Survey \citep{2013Msngr.154...35M}, and the VISIONS survey \citep{2023A&A...673A..58M}.} The left panel of Figure\,\ref{fig:motion} shows the direction of this motion (away from NGC\,2024 on the plane of the sky), alongside the proper motions of the confirmed YSOs considered by \citet{2026A&A...706A.210R} as NGC\,2024 members, and also the proper motions (wherever available) of any additional YSOs located within 8\arcmin{} of NGC\,2023. \textcolor{black}{The proper motion of J0541 is low,} consistent with the Orion star-forming complex; its youth also suggesting that it is not found far away from its birthplace. It is not included in the list of NGC\,2024 members from \citet{2026A&A...706A.210R}, but in proper-motion space, \textcolor{black}{J0541} falls inside the 1$\sigma$ covariance ellipse defined by those members (right panel of Figure\,\ref{fig:motion}). The Mahalanobis distance between \textcolor{black}{J0541} and the cluster centroid is \textcolor{black}{$D=0.48$}, suggesting a high degree of kinematic similarity to the NGC\,2024 population, \textcolor{black}{and supporting J0541 as a possible outer member of the cluster.} \textcolor{black}{The proper motions of YSOs located within 8\arcmin{} of NGC\,2023 are shown for comparison, but their membership status is yet to be assessed.} No proper-motion study of NGC\,2023 is currently available, highlighting the need for such an investigation.

\section{Summary} 
In this work, we searched for substellar objects in a region of \textcolor{black}{intermediate} and spatially variable reddening. We combined ERO data from Euclid (broadband photometry and spectroscopy) with archival observations from Spitzer (broadband photometry). We applied photometric criteria to compile a list of candidates, then we visually inspected their spectral traces to identify objects that exhibited substellar characteristics. We present a new spectroscopically confirmed substellar object of spectral type L3 and reddening $A_\mathrm{{V}}=3$\,mag, the first object of such a late spectral type to be discovered in this embedded area. This source lies close to the deuterium-burning mass limit, and could be of planetary mass, depending on its exact age.

The membership of \textcolor{black}{J0541} in the Orion star-forming complex is supported by its \textcolor{black}{location on the plane of the sky in proximity to NGC\,2023 (${\sim}$7\farcm5) and NGC\,2024 (${\sim} $20\arcmin{}), its near-infrared spectrum indicative of youth, its position in the CMDs in good agreement with theoretical isochrones, and its proper motion, which is consistent with the NGC\,2024 population. Its interstellar extinction further argues against an unreddened foreground origin.}

Multiple YSOs and dense cores have been discovered around NGC\,2023. This \textcolor{black}{star-forming region} serves as a potential laboratory for investigating how stellar feedback and local cloud structure shape the products of star formation. Now, for the first time, we are able to probe the substellar population of this dynamic environment, \textcolor{black}{with J0541 effectively pushing the low-mass limit of the population in this region.}

Objects of spectral type L3 or later have been identified in star-forming regions before \textcolor{black}{\cite[e.g.,][]{2012A&A...539A.151A}.} However, \textcolor{black}{young ultracool populations in star-forming environments affected by spatially variable extinction remain challenging to identify and characterize over large areas.} \textcolor{black}{J0541} is simultaneously of spectral type L3, young, close to the planetary-mass boundary, and affected by \textcolor{black}{intermediate} reddening. This combination \textcolor{black}{makes it a useful example of a young L-type object still associated with its natal cloud environment.}

Our discovery highlights Euclid's ability to reveal young \textcolor{black}{substellar objects in moderately obscured star-forming environments.} \textcolor{black}{J0541} therefore \textcolor{black}{becomes a reference} for the study of embedded very low-mass populations with wide-field, space-based surveys such as Euclid and the upcoming Nancy Grace Roman Space Telescope. \textcolor{black}{Such wide-field surveys are also highly complementary to the James Webb Space Telescope (JWST); while JWST can probe deeply embedded populations at high extinction, Euclid can extend searches for young, very low-mass objects over much larger areas at lower extinction. \textcolor{black}{J0541} thus represents a compelling target for JWST follow-up, illustrating the potential synergy between wide-field discovery with Euclid and detailed characterization with JWST.}

\begin{acknowledgments}
We thank I. Baraffe and M. Phillips for making available a digitized version of their ATMO isochrones in the Euclid passbands. Funding for S.T., M.Ž., C.D.T., E.L.M., N.S., S.M.T., N.V., K.B., and J.Y.Z. was provided by the European Union (ERC Advanced Grant, SUBSTELLAR, project No. 101054354). \textcolor{black}{N.L. acknowledges support from the Agencia  Estatal de Investigaci\'on del Ministerio de Ciencia e Innovaci\'on (AEI-MCINN) under grant PID2022-137241NB-C41.} J.Y.Z. thanks the support from the Western Postdoctoral Fellowship provided by Western University. \textcolor{black}{This work is part of grant CEX2025-001609-S, awarded to the Instituto de Astrofísica de Canarias under the Severo Ochoa Centre of Excellence program and funded by MICIU/AEI/10.13039/501100011033.} This work has made use of the Early Release Observations (ERO) data from the Euclid mission of the European Space Agency (ESA), 2024, doi:10.57780/esa-qmocze3. The authors acknowledge the Euclid Consortium, the European Space Agency, and a number of agencies and institutes that have supported the development of Euclid, in particular the Agenzia Spaziale Italiana, the Austrian Forschungsf\"{o}rderungsgesellschaft funded through BMK, the Belgian Science Policy, the Canadian Euclid Consortium, the Deutsches Zentrum f\"{u}r Luft- und Raumfahrt, the DTU Space and the Niels Bohr Institute in Denmark, the French Centre National d’Etudes Spatiales, the Fundaç\~{a}o para a Ci\^{e}ncia e a Tecnologia, the Hungarian Academy of Sciences, the Ministerio de Ciencia, Innovación y Universidades, the National Aeronautics and Space Administration, the National Astronomical Observatory of Japan, the Netherlandse Onderzoekschool Voor Astronomie, the Norwegian Space Agency, the Research Council of Finland, the Romanian Space Agency, the State Secretariat for Education, Research, and Innovation (SERI) at the Swiss Space Office (SSO), and the United Kingdom Space Agency. A complete and detailed list is available on the Euclid website (www.euclid-ec.org). This work is based in part on observations made with the Spitzer Space Telescope, which was operated by the Jet Propulsion Laboratory, California Institute of Technology, under a contract with NASA. This work is based in part on observations collected at the Centro Astronómico Hispano en Andalucía (CAHA) at Calar Alto, proposal 25B-2.2-023, operated jointly by Junta de Andalucía and Consejo Superior de Investigaciones Científicas (IAA-CSIC). \textcolor{black}{This work is based in part on data obtained as part of the UKIDSS project, defined in \citet{2007MNRAS.379.1599L}. UKIDSS uses the UKIRT Wide Field Camera \cite[WFCAM;][]{2007A&A...467..777C}. The photometric system is described in \citet{2006MNRAS.367..454H}, and the calibration is described in \citet{2009MNRAS.394..675H}. The pipeline processing and science archive are described in M. Irwin et al. (in preparation) and \citet{2008MNRAS.384..637H}. This work is based in part on observations made with VISTA \citep{2006SPIE.6269E..0XD, 2006Msngr.126...41E}.} This research has made use of the SIMBAD database \citep{2000A&AS..143....9W} operated at CDS, Strasbourg, France, and of the Astrophysics Data System, funded by NASA under Cooperative Agreement 80NSSC21M0056. This research has made use of the VizieR catalog access tool, CDS, Strasbourg, France \citep{10.26093/cds/vizier}. The original description 
of the VizieR service was published in \citet{vizier2000}. This research has made use of ``Aladin sky atlas'' \citep{2000A&AS..143...33B}, developed at CDS, Strasbourg Observatory, France. This publication makes use of VOSA \citep{2008A&A...492..277B}, developed under the Spanish Virtual Observatory (https://svo.cab.inta-csic.es) project funded by MCIN/AEI/10.13039/501100011033 through grant PID2023-146210NB-I00. VOSA has been partially updated using funding from the European Union's Horizon 2020 Research and Innovation Programme, under grant agreement No. 776403 (EXOPLANETS-A). This research has made use of the SVO Filter Profile Service ``Carlos Rodrigo'' \citep{2012ivoa.rept.1015R, 2020sea..confE.182R}, funded by MCIN/AEI/10.13039/501100011033/ through grant PID2023-146210NB-I00.
\end{acknowledgments}

\software{\texttt{astropy} (\!\!\citealt{2018AJ....156..123A}), \texttt{NumPy} \citep{2020Natur.585..357H}, \texttt{matplotlib} \citep{4160265}, \texttt{IPython} \citep{2007CSE.....9c..21P}, and \texttt{TOPCAT} \citep{2005ASPC..347...29T}.}

\bibliography{sample701}{}

@article{martinEuclidEarlyRelease2025,
  title = {{\emph{Euclid}}: {{Early Release Observations}} -- {{A}} Glance at Free-Floating Newborn Planets in the {$\sigma$} {{Orionis}} Cluster},
  shorttitle = {{\emph{Euclid}}},
  author = {Mart{\'i}n, E. L. and {\v Z}erjal, M. and Bouy, H. and {Martin-Gonzalez}, D. and Mu{\~n}oz Torres, S. and Barrado, D. and Olivares, J. and {P{\'e}rez-Garrido}, A. and {Mas-Buitrago}, P. and Cruz, P. and Solano, E. and Zapatero Osorio, M. R. and Lodieu, N. and B{\'e}jar, V. J. S. and Zhang, J.-Y. and Del Burgo, C. and Hu{\'e}lamo, N. and Laureijs, R. and Mora, A. and Saifollahi, T. and Cuillandre, J.-C. and Schirmer, M. and Tata, R. and Points, S. and {Phan-Bao}, N. and Goldman, B. and Casewell, S. L. and Reyl{\'e}, C. and Smart, R. L. and {Dominguez-Tagle}, C. and Escobar, A. and Sedighi, N. and Tsilia, S. and Vitas, N. and Ayadi, A. and Aghanim, N. and Altieri, B. and Andreon, S. and Auricchio, N. and Baldi, M. and Balestra, A. and Bardelli, S. and Basset, A. and Bender, R. and Bonino, D. and Branchini, E. and Brescia, M. and Brinchmann, J. and Camera, S. and Capobianco, V. and Carbone, C. and Carretero, J. and Casas, S. and Castellano, M. and Cavuoti, S. and Cimatti, A. and Congedo, G. and Conselice, C. J. and Conversi, L. and Copin, Y. and Corcione, L. and Courbin, F. and Courtois, H. M. and Cropper, M. and Da Silva, A. and Degaudenzi, H. and Di Giorgio, A. M. and Dinis, J. and Dubath, F. and Dupac, X. and Dusini, S. and Ealet, A. and Farina, M. and Farrens, S. and Ferriol, S. and Fosalba, P. and Frailis, M. and Franceschi, E. and Fumana, M. and Galeotta, S. and Garilli, B. and Gillard, W. and Gillis, B. and Giocoli, C. and {G{\'o}mez-Alvarez}, P. and Grazian, A. and Grupp, F. and Guzzo, L. and Haugan, S. V. H. and Hoar, J. and Hoekstra, H. and Holmes, W. and Hook, I. and Hormuth, F. and Hornstrup, A. and Hu, D. and Hudelot, P. and Jahnke, K. and Jhabvala, M. and Keih{\"a}nen, E. and Kermiche, S. and Kiessling, A. and Kilbinger, M. and Kitching, T. and Kohley, R. and Kubik, B. and K{\"u}mmel, M. and Kunz, M. and {Kurki-Suonio}, H. and Le Mignant, D. and Ligori, S. and Lilje, P. B. and Lindholm, V. and Lloro, I. and Maino, D. and Maiorano, E. and Mansutti, O. and Marggraf, O. and Martinet, N. and Marulli, F. and Massey, R. and Medinaceli, E. and Mei, S. and Melchior, M. and Mellier, Y. and Meneghetti, M. and Meylan, G. and Mohr, J. J. and Moresco, M. and Moscardini, L. and Niemi, S.-M. and Padilla, C. and Paltani, S. and Pasian, F. and Pedersen, K. and Percival, W. J. and Pettorino, V. and Pires, S. and Polenta, G. and Poncet, M. and Popa, L. A. and Pozzetti, L. and Racca, G. D. and Raison, F. and Rebolo, R. and Renzi, A. and Rhodes, J. and Riccio, G. and Rix, Hans-Walter and Romelli, E. and Roncarelli, M. and Rossetti, E. and Saglia, R. and Sapone, D. and Sartoris, B. and Sauvage, M. and Scaramella, R. and Schneider, P. and Secroun, A. and Seidel, G. and Seiffert, M. and Serrano, S. and Sirignano, C. and Sirri, G. and Stanco, L. and {Tallada-Cresp{\'i}}, P. and Taylor, A. N. and Teplitz, H. I. and Tereno, I. and {Toledo-Moreo}, R. and Tsyganov, A. and Tutusaus, I. and Valenziano, L. and Vassallo, T. and Verdoes Kleijn, G. and Wang, Y. and Weller, J. and Williams, O. R. and Zucca, E. and Baccigalupi, C. and Willis, G. and Simon, P. and {Mart{\'i}n-Fleitas}, J. and Scott, D.},
  year = 2025,
  month = may,
  journal = {A\&A},
  volume = {697},
  pages = {A7},
  publisher = {EDP Sciences},
  issn = {0004-6361, 1432-0746},
  doi = {10.1051/0004-6361/202450793},
  urldate = {2025-07-14},
  copyright = {https://creativecommons.org/licenses/by/4.0},
  langid = {english}
}

@article{penaramirezNEWISOLATEDPLANETARYMASS2012,
  title = {{{NEW ISOLATED PLANETARY-MASS OBJECTS AND THE STELLAR AND SUBSTELLAR MASS FUNCTION OF THE}} {$\sigma$} {{ORIONIS CLUSTER}}},
  author = {Pe{\~n}a Ram{\'i}rez, K. and B{\'e}jar, V. J. S. and Zapatero Osorio, M. R. and {Petr-Gotzens}, M. G. and Mart{\'i}n, E. L.},
  year = 2012,
  month = jul,
  journal = {ApJ},
  volume = {754},
  number = {1},
  pages = {30},
  issn = {0004-637X, 1538-4357},
  doi = {10.1088/0004-637X/754/1/30},
  urldate = {2025-04-06},
  langid = {english}
}

@ARTICLE{2020A&A...637A..38P,
       author = {{Phillips}, M.~W. and {Tremblin}, P. and {Baraffe}, I. and {Chabrier}, G. and {Allard}, N.~F. and {Spiegelman}, F. and {Goyal}, J.~M. and {Drummond}, B. and {H{\'e}brard}, E.},
        title = "{A new set of atmosphere and evolution models for cool T-Y brown dwarfs and giant exoplanets}",
      journal = {\aap},
         year = 2020,
        month = may,
       volume = {637},
          eid = {A38},
        pages = {A38},
          doi = {10.1051/0004-6361/201937381},
archivePrefix = {arXiv},
       eprint = {2003.13717},
 primaryClass = {astro-ph.SR},
       adsurl = {https://ui.adsabs.harvard.edu/abs/2020A&A...637A..38P}
}

@ARTICLE{2013MNRAS.429..775L,
       author = {{L{\'o}pez-Garc{\'\i}a}, M.~A. and {L{\'o}pez-Santiago}, J. and {Albacete-Colombo}, J.~F. and {P{\'e}rez-Gonz{\'a}lez}, P.~G. and {de Castro}, E.},
        title = "{A comprehensive study of NGC 2023 with XMM-Newton and Spitzer}",
      journal = {\mnras},
         year = 2013,
        month = feb,
       volume = {429},
       number = {1},
        pages = {775-791},
          doi = {10.1093/mnras/sts373},
archivePrefix = {arXiv},
       eprint = {1211.2935},
 primaryClass = {astro-ph.SR},
       adsurl = {https://ui.adsabs.harvard.edu/abs/2013MNRAS.429..775L}
}

@ARTICLE{2025A&A...697A...6C,
       author = {{Cuillandre}, J.-C. and {Bertin}, E. and {Bolzonella}, M. and {Bouy}, H. and {Gwyn}, S. and {Isani}, S. and {Kluge}, M. and {Lai}, O. and {Lan{\c{c}}on}, A. and {Lang}, D.~A. and {Laureijs}, R. and {Saifollahi}, T. and {Schirmer}, M. and {Stone}, C. and {Abdurro'uf} and {Aghanim}, N. and {Altieri}, B. and {Annibali}, F. and {Atek}, H. and {Awad}, P. and {Baes}, M. and {Ba{\~n}ados}, E. and {Barrado}, D. and {Belladitta}, S. and {Belokurov}, V. and {Boselli}, A. and {Bournaud}, F. and {Bovy}, J. and {Bowler}, R.~A.~A. and {Buenadicha}, G. and {Buitrago}, F. and {Cantiello}, M. and {Carollo}, D. and {Codis}, S. and {Collins}, M.~L.~M. and {Congedo}, G. and {Dalessandro}, E. and {de Lapparent}, V. and {De Paolis}, F. and {Diego}, J.~M. and {Dimauro}, P. and {Dinis}, J. and {Dole}, H. and {Duc}, P.-A. and {Erkal}, D. and {Ezziati}, M. and {Ferguson}, A.~M.~N. and {Ferr{\'e}-Mateu}, A. and {Franco}, A. and {Gavazzi}, R. and {George}, K. and {Gillard}, W. and {Golden-Marx}, J.~B. and {Goldman}, B. and {Gonzalez}, A.~H. and {Habas}, R. and {Hartley}, W.~G. and {Hatch}, N.~A. and {Kohley}, R. and {Hoar}, J. and {Howell}, J.~M. and {Hunt}, L.~K. and {Jablonka}, P. and {Jauzac}, M. and {Kang}, Y. and {Knapen}, J.~H. and {Kneib}, J.-P. and {Kuzma}, P.~B. and {Larsen}, S.~S. and {Marchal}, O. and {Mart{\'\i}n-Fleitas}, J. and {Marcos-Arenal}, P. and {Marleau}, F.~R. and {Mart{\'\i}n}, E.~L. and {Massari}, D. and {McConnachie}, A.~W. and {Meneghetti}, M. and {Miluzio}, M. and {Miro Carretero}, J. and {Miyatake}, H. and {Mondelin}, M. and {Montes}, M. and {Mora}, A. and {M{\"u}ller}, O. and {Nally}, C. and {Noeske}, K. and {Nucita}, A.~A. and {Oesch}, P.~A. and {Oguri}, M. and {Peletier}, R.~F. and {Poulain}, M. and {Quilley}, L. and {Racca}, G.~D. and {Rejkuba}, M. and {Rhodes}, J. and {Rocci}, P.-F. and {Rom{\'a}n}, J. and {Sacquegna}, S. and {Saremi}, E. and {Scaramella}, R. and {Schinnerer}, E. and {Serjeant}, S. and {Sola}, E. and {Sorce}, J.~G. and {Tarsitano}, F. and {Tereno}, I. and {Toft}, S. and {Tortora}, C. and {Urbano}, M. and {Venhola}, A. and {Voggel}, K. and {Weaver}, J.~R. and {Xu}, X. and {{\v{Z}}erjal}, M. and {Z{\"o}ller}, R. and {Andreon}, S. and {Auricchio}, N. and {Baccigalupi}, C. and {Baldi}, M. and {Balestra}, A. and {Bardelli}, S. and {Basset}, A. and {Bender}, R. and {Bodendorf}, C. and {Branchini}, E. and {Brau-Nogue}, S. and {Brescia}, M. and {Brinchmann}, J. and {Camera}, S. and {Capobianco}, V. and {Carbone}, C. and {Carretero}, J. and {Casas}, S. and {Castander}, F.~J. and {Castellano}, M. and {Cavuoti}, S. and {Cimatti}, A. and {Conselice}, C.~J. and {Conversi}, L. and {Copin}, Y. and {Courbin}, F. and {Courtois}, H.~M. and {Cropper}, M. and {Cuby}, J.-G. and {Da Silva}, A. and {Degaudenzi}, H. and {Di Giorgio}, A.~M. and {Douspis}, M. and {Duncan}, C.~A.~J. and {Dupac}, X. and {Dusini}, S. and {Fabricius}, M. and {Farina}, M. and {Farrens}, S. and {Ferriol}, S. and {Fosalba}, P. and {Fotopoulou}, S. and {Frailis}, M. and {Franceschi}, E. and {Galeotta}, S. and {Garilli}, B. and {Gillis}, B. and {Giocoli}, C. and {G{\'o}mez-Alvarez}, P. and {Grazian}, A. and {Grupp}, F. and {Guzzo}, L. and {Haugan}, S.~V.~H. and {Hoekstra}, H. and {Holmes}, W. and {Hook}, I. and {Hormuth}, F. and {Hornstrup}, A. and {Hudelot}, P. and {Jahnke}, K. and {Jhabvala}, M. and {Keih{\"a}nen}, E. and {Kermiche}, S. and {Kiessling}, A. and {Kilbinger}, M. and {Kitching}, T. and {Kubik}, B. and {Kuijken}, K. and {K{\"u}mmel}, M. and {Kunz}, M. and {Kurki-Suonio}, H. and {Lahav}, O. and {Liebing}, P. and {Ligori}, S. and {Lilje}, P.~B. and {Lindholm}, V. and {Lloro}, I. and {Maino}, D. and {Maiorano}, E. and {Mansutti}, O. and {Marggraf}, O. and {Markovic}, K. and {Martinet}, N. and {Marulli}, F. and {Massey}, R.},
        title = "{Euclid: Early Release Observations {\textendash} Programme overview and pipeline for compact- and diffuse-emission photometry}",
      journal = {\aap},
         year = 2025,
        month = may,
       volume = {697},
          eid = {A6},
        pages = {A6},
          doi = {10.1051/0004-6361/202450803},
archivePrefix = {arXiv},
       eprint = {2405.13496},
 primaryClass = {astro-ph.IM},
       adsurl = {https://ui.adsabs.harvard.edu/abs/2025A&A...697A...6C}
}

@ARTICLE{1963ApJ...137.1121K,
       author = {{Kumar}, Shiv S.},
        title = "{The Structure of Stars of Very Low Mass.}",
      journal = {\apj},
         year = 1963,
        month = may,
       volume = {137},
        pages = {1121},
          doi = {10.1086/147589},
       adsurl = {https://ui.adsabs.harvard.edu/abs/1963ApJ...137.1121K}
}

@ARTICLE{1963PThPh..30..460H,
       author = {{Hayashi}, C. and {Nakano}, T.},
        title = "{Evolution of Stars of Small Masses in the Pre-Main-Sequence Stages}",
      journal = {Progress of Theoretical Physics},
         year = 1963,
        month = oct,
       volume = {30},
       number = {4},
        pages = {460-474},
          doi = {10.1143/PTP.30.460},
       adsurl = {https://ui.adsabs.harvard.edu/abs/1963PThPh..30..460H}
}

@ARTICLE{2021ApJS..253....7K,
       author = {{Kirkpatrick}, J. Davy and {Gelino}, Christopher R. and {Faherty}, Jacqueline K. and {Meisner}, Aaron M. and {Caselden}, Dan and {Schneider}, Adam C. and {Marocco}, Federico and {Cayago}, Alfred J. and {Smart}, R.~L. and {Eisenhardt}, Peter R. and {Kuchner}, Marc J. and {Wright}, Edward L. and {Cushing}, Michael C. and {Allers}, Katelyn N. and {Bardalez Gagliuffi}, Daniella C. and {Burgasser}, Adam J. and {Gagn{\'e}}, Jonathan and {Logsdon}, Sarah E. and {Martin}, Emily C. and {Ingalls}, James G. and {Lowrance}, Patrick J. and {Abrahams}, Ellianna S. and {Aganze}, Christian and {Gerasimov}, Roman and {Gonzales}, Eileen C. and {Hsu}, Chih-Chun and {Kamraj}, Nikita and {Kiman}, Rocio and {Rees}, Jon and {Theissen}, Christopher and {Ammar}, Kareem and {Andersen}, Nikolaj Stevnbak and {Beaulieu}, Paul and {Colin}, Guillaume and {Elachi}, Charles A. and {Goodman}, Samuel J. and {Gramaize}, L{\'e}opold and {Hamlet}, Leslie K. and {Hong}, Justin and {Jonkeren}, Alexander and {Khalil}, Mohammed and {Martin}, David W. and {Pendrill}, William and {Pumphrey}, Benjamin and {Rothermich}, Austin and {Sainio}, Arttu and {Stenner}, Andres and {Tanner}, Christopher and {Th{\'e}venot}, Melina and {Voloshin}, Nikita V. and {Walla}, Jim and {W{\k{e}}dracki}, Zbigniew and {Backyard Worlds: Planet 9 Collaboration}},
        title = "{The Field Substellar Mass Function Based on the Full-sky 20 pc Census of 525 L, T, and Y Dwarfs}",
      journal = {\apjs},
         year = 2021,
        month = mar,
       volume = {253},
       number = {1},
          eid = {7},
        pages = {7},
          doi = {10.3847/1538-4365/abd107},
archivePrefix = {arXiv},
       eprint = {2011.11616},
 primaryClass = {astro-ph.SR},
       adsurl = {https://ui.adsabs.harvard.edu/abs/2021ApJS..253....7K}
}

@ARTICLE{2021A&A...655A..44E,
       author = {{Euclid Collaboration} and {Pocino}, A. and {Tutusaus}, I. and {Castander}, F.~J. and {Fosalba}, P. and {Crocce}, M. and {Porredon}, A. and {Camera}, S. and {Cardone}, V. and {Casas}, S. and {Kitching}, T. and {Lacasa}, F. and {Martinelli}, M. and {Pourtsidou}, A. and {Sakr}, Z. and {Andreon}, S. and {Auricchio}, N. and {Baccigalupi}, C. and {Balaguera-Antol{\'\i}nez}, A. and {Baldi}, M. and {Balestra}, A. and {Bardelli}, S. and {Bender}, R. and {Biviano}, A. and {Bodendorf}, C. and {Bonino}, D. and {Boucaud}, A. and {Bozzo}, E. and {Branchini}, E. and {Brescia}, M. and {Brinchmann}, J. and {Burigana}, C. and {Cabanac}, R. and {Capobianco}, V. and {Cappi}, A. and {Carvalho}, C.~S. and {Castellano}, M. and {Castignani}, G. and {Cavuoti}, S. and {Cimatti}, A. and {Cledassou}, R. and {Colodro-Conde}, C. and {Congedo}, G. and {Conselice}, C.~J. and {Conversi}, L. and {Copin}, Y. and {Corcione}, L. and {Costille}, A. and {Coupon}, J. and {Courtois}, H.~M. and {Cropper}, M. and {Cuby}, J.-G. and {Da Silva}, A. and {de la Torre}, S. and {Di Ferdinando}, D. and {Dubath}, F. and {Duncan}, C. and {Dupac}, X. and {Dusini}, S. and {Farrens}, S. and {Ferreira}, P.~G. and {Ferrero}, I. and {Finelli}, F. and {Fotopoulou}, S. and {Frailis}, M. and {Franceschi}, E. and {Galeotta}, S. and {Garilli}, B. and {Gillard}, W. and {Gillis}, B. and {Giocoli}, C. and {Gozaliasl}, G. and {Graci{\'a}-Carpio}, J. and {Grupp}, F. and {Guzzo}, L. and {Holmes}, W. and {Hormuth}, F. and {Jahnke}, K. and {Keihanen}, E. and {Kermiche}, S. and {Kiessling}, A. and {Kirkpatrick}, C.~C. and {Kunz}, M. and {Kurki-Suonio}, H. and {Ligori}, S. and {Lilje}, P.~B. and {Lloro}, I. and {Maino}, D. and {Maiorano}, E. and {Mansutti}, O. and {Marggraf}, O. and {Martinet}, N. and {Marulli}, F. and {Massey}, R. and {Maurogordato}, S. and {Medinaceli}, E. and {Mei}, S. and {Meneghetti}, M. and {Benton Metcalf}, R. and {Meylan}, G. and {Moresco}, M. and {Morin}, B. and {Moscardini}, L. and {Munari}, E. and {Nakajima}, R. and {Neissner}, C. and {Nichol}, R.~C. and {Niemi}, S. and {Nightingale}, J. and {Padilla}, C. and {Paltani}, S. and {Pasian}, F. and {Patrizii}, L. and {Pedersen}, K. and {Percival}, W.~J. and {Pettorino}, V. and {Pires}, S. and {Polenta}, G. and {Poncet}, M. and {Popa}, L. and {Potter}, D. and {Pozzetti}, L. and {Raison}, F. and {Renzi}, A. and {Rhodes}, J. and {Riccio}, G. and {Romelli}, E. and {Roncarelli}, M. and {Rossetti}, E. and {Saglia}, R. and {S{\'a}nchez}, A.~G. and {Sapone}, D. and {Scaramella}, R. and {Schneider}, P. and {Scottez}, V. and {Secroun}, A. and {Seidel}, G. and {Serrano}, S. and {Sirignano}, C. and {Sirri}, G. and {Stanco}, L. and {Sureau}, F. and {Taylor}, A.~N. and {Tenti}, M. and {Tereno}, I. and {Teyssier}, R. and {Toledo-Moreo}, R. and {Tramacere}, A. and {Valentijn}, E.~A. and {Valenziano}, L. and {Valiviita}, J. and {Vassallo}, T. and {Viel}, M. and {Wang}, Y. and {Welikala}, N. and {Whittaker}, L. and {Zacchei}, A. and {Zamorani}, G. and {Zoubian}, J. and {Zucca}, E.},
        title = "{Euclid preparation. XII. Optimizing the photometric sample of the Euclid survey for galaxy clustering and galaxy-galaxy lensing analyses}",
      journal = {\aap},
         year = 2021,
        month = nov,
       volume = {655},
          eid = {A44},
        pages = {A44},
          doi = {10.1051/0004-6361/202141061},
archivePrefix = {arXiv},
       eprint = {2104.05698},
 primaryClass = {astro-ph.CO},
       adsurl = {https://ui.adsabs.harvard.edu/abs/2021A&A...655A..44E}
}

@ARTICLE{2022A&A...662A.112E,
       author = {{Euclid Collaboration} and {Scaramella}, R. and {Amiaux}, J. and {Mellier}, Y. and {Burigana}, C. and {Carvalho}, C.~S. and {Cuillandre}, J.-C. and {Da Silva}, A. and {Derosa}, A. and {Dinis}, J. and {Maiorano}, E. and {Maris}, M. and {Tereno}, I. and {Laureijs}, R. and {Boenke}, T. and {Buenadicha}, G. and {Dupac}, X. and {Gaspar Venancio}, L.~M. and {G{\'o}mez-{\'A}lvarez}, P. and {Hoar}, J. and {Lorenzo Alvarez}, J. and {Racca}, G.~D. and {Saavedra-Criado}, G. and {Schwartz}, J. and {Vavrek}, R. and {Schirmer}, M. and {Aussel}, H. and {Azzollini}, R. and {Cardone}, V.~F. and {Cropper}, M. and {Ealet}, A. and {Garilli}, B. and {Gillard}, W. and {Granett}, B.~R. and {Guzzo}, L. and {Hoekstra}, H. and {Jahnke}, K. and {Kitching}, T. and {Maciaszek}, T. and {Meneghetti}, M. and {Miller}, L. and {Nakajima}, R. and {Niemi}, S.~M. and {Pasian}, F. and {Percival}, W.~J. and {Pottinger}, S. and {Sauvage}, M. and {Scodeggio}, M. and {Wachter}, S. and {Zacchei}, A. and {Aghanim}, N. and {Amara}, A. and {Auphan}, T. and {Auricchio}, N. and {Awan}, S. and {Balestra}, A. and {Bender}, R. and {Bodendorf}, C. and {Bonino}, D. and {Branchini}, E. and {Brau-Nogue}, S. and {Brescia}, M. and {Candini}, G.~P. and {Capobianco}, V. and {Carbone}, C. and {Carlberg}, R.~G. and {Carretero}, J. and {Casas}, R. and {Castander}, F.~J. and {Castellano}, M. and {Cavuoti}, S. and {Cimatti}, A. and {Cledassou}, R. and {Congedo}, G. and {Conselice}, C.~J. and {Conversi}, L. and {Copin}, Y. and {Corcione}, L. and {Costille}, A. and {Courbin}, F. and {Degaudenzi}, H. and {Douspis}, M. and {Dubath}, F. and {Duncan}, C.~A.~J. and {Dusini}, S. and {Farrens}, S. and {Ferriol}, S. and {Fosalba}, P. and {Fourmanoit}, N. and {Frailis}, M. and {Franceschi}, E. and {Franzetti}, P. and {Fumana}, M. and {Gillis}, B. and {Giocoli}, C. and {Grazian}, A. and {Grupp}, F. and {Haugan}, S.~V.~H. and {Holmes}, W. and {Hormuth}, F. and {Hudelot}, P. and {Kermiche}, S. and {Kiessling}, A. and {Kilbinger}, M. and {Kohley}, R. and {Kubik}, B. and {K{\"u}mmel}, M. and {Kunz}, M. and {Kurki-Suonio}, H. and {Lahav}, O. and {Ligori}, S. and {Lilje}, P.~B. and {Lloro}, I. and {Mansutti}, O. and {Marggraf}, O. and {Markovic}, K. and {Marulli}, F. and {Massey}, R. and {Maurogordato}, S. and {Melchior}, M. and {Merlin}, E. and {Meylan}, G. and {Mohr}, J.~J. and {Moresco}, M. and {Morin}, B. and {Moscardini}, L. and {Munari}, E. and {Nichol}, R.~C. and {Padilla}, C. and {Paltani}, S. and {Peacock}, J. and {Pedersen}, K. and {Pettorino}, V. and {Pires}, S. and {Poncet}, M. and {Popa}, L. and {Pozzetti}, L. and {Raison}, F. and {Rebolo}, R. and {Rhodes}, J. and {Rix}, H.-W. and {Roncarelli}, M. and {Rossetti}, E. and {Saglia}, R. and {Schneider}, P. and {Schrabback}, T. and {Secroun}, A. and {Seidel}, G. and {Serrano}, S. and {Sirignano}, C. and {Sirri}, G. and {Skottfelt}, J. and {Stanco}, L. and {Starck}, J.~L. and {Tallada-Cresp{\'\i}}, P. and {Tavagnacco}, D. and {Taylor}, A.~N. and {Teplitz}, H.~I. and {Toledo-Moreo}, R. and {Torradeflot}, F. and {Trifoglio}, M. and {Valentijn}, E.~A. and {Valenziano}, L. and {Verdoes Kleijn}, G.~A. and {Wang}, Y. and {Welikala}, N. and {Weller}, J. and {Wetzstein}, M. and {Zamorani}, G. and {Zoubian}, J. and {Andreon}, S. and {Baldi}, M. and {Bardelli}, S. and {Boucaud}, A. and {Camera}, S. and {Di Ferdinando}, D. and {Fabbian}, G. and {Farinelli}, R. and {Galeotta}, S. and {Graci{\'a}-Carpio}, J. and {Maino}, D. and {Medinaceli}, E. and {Mei}, S. and {Neissner}, C. and {Polenta}, G. and {Renzi}, A. and {Romelli}, E. and {Rosset}, C. and {Sureau}, F. and {Tenti}, M. and {Vassallo}, T. and {Zucca}, E. and {Baccigalupi}, C. and {Balaguera-Antol{\'\i}nez}, A. and {Battaglia}, P. and {Biviano}, A. and {Borgani}, S. and {Bozzo}, E. and {Cabanac}, R. and {Cappi}, A.},
        title = "{Euclid preparation. I. The Euclid Wide Survey}",
      journal = {\aap},
         year = 2022,
        month = jun,
       volume = {662},
          eid = {A112},
        pages = {A112},
          doi = {10.1051/0004-6361/202141938},
archivePrefix = {arXiv},
       eprint = {2108.01201},
 primaryClass = {astro-ph.CO},
       adsurl = {https://ui.adsabs.harvard.edu/abs/2022A&A...662A.112E}
}

@ARTICLE{2025A&A...697A...2E,
       author = {{Euclid Collaboration} and {Cropper}, M.~S. and {Al-Bahlawan}, A. and {Amiaux}, J. and {Awan}, S. and {Azzollini}, R. and {Benson}, K. and {Berthe}, M. and {Boucher}, J. and {Bozzo}, E. and {Brockley-Blatt}, C. and {Candini}, G.~P. and {Cara}, C. and {Chaudery}, R.~A. and {Cole}, R.~E. and {Danto}, P. and {Denniston}, J. and {Di Giorgio}, A.~M. and {Dryer}, B. and {Dubois}, J.-P. and {Endicott}, J. and {Farina}, M. and {Galli}, E. and {Genolet}, L. and {Gow}, J.~P.~D. and {Guttridge}, P. and {Hailey}, M. and {Hall}, D. and {Harper}, C. and {Hoekstra}, H. and {Holland}, A.~D. and {Horeau}, B. and {Hu}, D. and {James}, R.~E. and {Khalil}, A. and {King}, R. and {Kitching}, T. and {Kohley}, R. and {Larcheveque}, C. and {Lawrenson}, A. and {Liebing}, P. and {Liu}, S.~J. and {Martignac}, J. and {Massey}, R. and {McCracken}, H.~J. and {Miller}, L. and {Murray}, N. and {Nakajima}, R. and {Niemi}, S.-M. and {Nightingale}, J.~W. and {Paltani}, S. and {Pendem}, A. and {Philippon}, A. and {Plana}, C. and {Pool}, P. and {Pottinger}, S. and {Racca}, G.~D. and {Rhodes}, J. and {Rousseau}, A. and {Ruane}, K. and {Salatti}, M. and {Salvignol}, J.-C. and {Sciortino}, A. and {Short}, A. and {Skottfelt}, J. and {Smit}, S.~J.~A. and {Swindells}, I. and {Szafraniec}, M. and {Thomas}, P.~D. and {Thomas}, W. and {Tommasi}, E. and {Tosti}, S. and {Visticot}, F. and {Walton}, D.~M. and {Willis}, G. and {Winter}, B. and {Aghanim}, N. and {Altieri}, B. and {Amara}, A. and {Andreon}, S. and {Auricchio}, N. and {Aussel}, H. and {Baccigalupi}, C. and {Baldi}, M. and {Balestra}, A. and {Bardelli}, S. and {Basset}, A. and {Bender}, R. and {Bernardeau}, F. and {Bodendorf}, C. and {Boenke}, T. and {Bonino}, D. and {Branchini}, E. and {Brescia}, M. and {Brinchmann}, J. and {Camera}, S. and {Capobianco}, V. and {Carbone}, C. and {Cardone}, V.~F. and {Carretero}, J. and {Casas}, R. and {Casas}, S. and {Castander}, F.~J. and {Castellano}, M. and {Castignani}, G. and {Cavuoti}, S. and {Cimatti}, A. and {Colodro-Conde}, C. and {Congedo}, G. and {Conselice}, C.~J. and {Conversi}, L. and {Copin}, Y. and {Courbin}, F. and {Courtois}, H.~M. and {Crocce}, M. and {Cuby}, J.-G. and {Cuillandre}, J.-C. and {Da Silva}, A. and {Degaudenzi}, H. and {De Lucia}, G. and {Dinis}, J. and {Dolding}, C. and {Douspis}, M. and {Duncan}, C.~A.~J. and {Dupac}, X. and {Dusini}, S. and {Ealet}, A. and {Fabricius}, M. and {Farrens}, S. and {Ferriol}, S. and {Fosalba}, P. and {Fotopoulou}, S. and {Frailis}, M. and {Franceschi}, E. and {Franzetti}, P. and {Frugier}, P.-A. and {Fumana}, M. and {Galeotta}, S. and {Garilli}, B. and {George}, K. and {Gillard}, W. and {Gillis}, B. and {Giocoli}, C. and {G{\'o}mez-Alvarez}, P. and {Granett}, B.~R. and {Grazian}, A. and {Grupp}, F. and {Guzzo}, L. and {Haugan}, S.~V.~H. and {Herent}, O. and {Hoar}, J. and {Holliman}, M.~S. and {Holmes}, W. and {Hook}, I. and {Hormuth}, F. and {Hornstrup}, A. and {Hudelot}, P. and {Ili{\'c}}, S. and {Jahnke}, K. and {Jhabvala}, M. and {Joachimi}, B. and {Keih{\"a}nen}, E. and {Kermiche}, S. and {Kiessling}, A. and {Kilbinger}, M. and {Kubik}, B. and {Kuijken}, K. and {K{\"u}mmel}, M. and {Kunz}, M. and {Kurki-Suonio}, H. and {Lahav}, O. and {Laureijs}, R. and {Ligori}, S. and {Lilje}, P.~B. and {Lindholm}, V. and {Lloro}, I. and {Lorenzo Alvarez}, J. and {Mainetti}, G. and {Maino}, D. and {Maiorano}, E. and {Mansutti}, O. and {Marcin}, S. and {Marggraf}, O. and {Markovic}, K. and {Martinelli}, M. and {Martinet}, N. and {Marulli}, F. and {Masters}, D.~C. and {Maurogordato}, S. and {Medinaceli}, E. and {Mei}, S. and {Melchior}, M. and {Mellier}, Y. and {Meneghetti}, M. and {Merlin}, E. and {Meylan}, G. and {Mohr}, J.~J. and {Moresco}, M. and {Moscardini}, L. and {Neissner}, C.},
        title = "{Euclid: II. The VIS instrument}",
      journal = {\aap},
         year = 2025,
        month = may,
       volume = {697},
          eid = {A2},
        pages = {A2},
          doi = {10.1051/0004-6361/202450996},
archivePrefix = {arXiv},
       eprint = {2405.13492},
 primaryClass = {astro-ph.IM},
       adsurl = {https://ui.adsabs.harvard.edu/abs/2025A&A...697A...2E}
}

@ARTICLE{2025A&A...697A...3E,
       author = {{Euclid Collaboration} and {Jahnke}, K. and {Gillard}, W. and {Schirmer}, M. and {Ealet}, A. and {Maciaszek}, T. and {Prieto}, E. and {Barbier}, R. and {Bonoli}, C. and {Corcione}, L. and {Dusini}, S. and {Grupp}, F. and {Hormuth}, F. and {Ligori}, S. and {Martin}, L. and {Morgante}, G. and {Padilla}, C. and {Toledo-Moreo}, R. and {Trifoglio}, M. and {Valenziano}, L. and {Bender}, R. and {Castander}, F.~J. and {Garilli}, B. and {Lilje}, P.~B. and {Rix}, H.-W. and {Andersen}, M.~I. and {Auricchio}, N. and {Balestra}, A. and {Barriere}, J.-C. and {Battaglia}, P. and {Berthe}, M. and {Bodendorf}, C. and {Boenke}, T. and {Bon}, W. and {Bonnefoi}, A. and {Caillat}, A. and {Capobianco}, V. and {Carle}, M. and {Casas}, R. and {Cho}, H. and {Costille}, A. and {Ducret}, F. and {Ferriol}, S. and {Franceschi}, E. and {Gimenez}, J.-L. and {Holmes}, W. and {Hornstrup}, A. and {Jhabvala}, M. and {Kohley}, R. and {Kubik}, B. and {Laureijs}, R. and {Le Mignant}, D. and {Lloro}, I. and {Medinaceli}, E. and {Mellier}, Y. and {Polenta}, G. and {Racca}, G.~D. and {Renzi}, A. and {Salvignol}, J.-C. and {Secroun}, A. and {Seidel}, G. and {Seiffert}, M. and {Sirignano}, C. and {Sirri}, G. and {Strada}, P. and {Smadja}, G. and {Stanco}, L. and {Wachter}, S. and {Anselmi}, S. and {Borsato}, E. and {Caillat}, L. and {Cogato}, F. and {Colodro-Conde}, C. and {Crouzet}, P.-E. and {Conforti}, V. and {D'Alessandro}, M. and {Copin}, Y. and {Cuillandre}, J.-C. and {Davies}, J.~E. and {Davini}, S. and {Derosa}, A. and {Diaz}, J.~J. and {Di Domizio}, S. and {Di Ferdinando}, D. and {Farinelli}, R. and {Ferrari}, A.~G. and {Fornari}, F. and {Gabarra}, L. and {Garcia}, R. and {Gutierrez}, C.~M. and {Giacomini}, F. and {Lagier}, P. and {Gianotti}, F. and {Krause}, O. and {Madrid}, F. and {Laudisio}, F. and {Macias-Perez}, J. and {Naletto}, G. and {Niclas}, M. and {Marpaud}, J. and {Mauri}, N. and {da Silva}, R. and {Passalacqua}, F. and {Paterson}, K. and {Patrizii}, L. and {Risso}, I. and {Solheim}, B.~G.~B. and {Scodeggio}, M. and {Stassi}, P. and {Steinwagner}, J. and {Tenti}, M. and {Testera}, G. and {Travaglini}, R. and {Tosi}, S. and {Troja}, A. and {Tubio}, O. and {Valieri}, C. and {Vescovi}, C. and {Ventura}, S. and {Aghanim}, N. and {Altieri}, B. and {Amara}, A. and {Amiaux}, J. and {Andreon}, S. and {Appleton}, P.~N. and {Aussel}, H. and {Baccigalupi}, C. and {Baldi}, M. and {Bardelli}, S. and {Basset}, A. and {Bonchi}, A. and {Bonino}, D. and {Branchini}, E. and {Brescia}, M. and {Brinchmann}, J. and {Camera}, S. and {Carbone}, C. and {Cardone}, V.~F. and {Carretero}, J. and {Casas}, S. and {Castellano}, M. and {Castignani}, G. and {Cavuoti}, S. and {Chabaud}, P.-Y. and {Cimatti}, A. and {Congedo}, G. and {Conselice}, C.~J. and {Conversi}, L. and {Courbin}, F. and {Courtois}, H.~M. and {Crocce}, M. and {Cropper}, M. and {Cuby}, J.-G. and {Da Silva}, A. and {Degaudenzi}, H. and {De Lucia}, G. and {Di Giorgio}, A.~M. and {Dinis}, J. and {Douspis}, M. and {Dubath}, F. and {Duncan}, C.~A.~J. and {Dupac}, X. and {Fabricius}, M. and {Farina}, M. and {Farrens}, S. and {Faustini}, F. and {Fosalba}, P. and {Fotopoulou}, S. and {Fourmanoit}, N. and {Frailis}, M. and {Franzetti}, P. and {Galeotta}, S. and {George}, K. and {Gillis}, B. and {Giocoli}, C. and {G{\'o}mez-Alvarez}, P. and {Granett}, B.~R. and {Grazian}, A. and {Guzzo}, L. and {Hailey}, M. and {Haugan}, S.~V.~H. and {Hoar}, J. and {Hoekstra}, H. and {Hook}, I. and {Hudelot}, P. and {Ili{\'c}}, S. and {Joachimi}, B. and {Keih{\"a}nen}, E. and {Kermiche}, S. and {Kiessling}, A. and {Kilbinger}, M. and {Kitching}, T. and {K{\"u}mmel}, M. and {Kunz}, M. and {Kurki-Suonio}, H. and {Lahav}, O. and {Liebing}, P. and {Lindholm}, V. and {Lorenzo Alvarez}, J. and {Mainetti}, G.},
        title = "{Euclid: III. The NISP Instrument}",
      journal = {\aap},
         year = 2025,
        month = may,
       volume = {697},
          eid = {A3},
        pages = {A3},
          doi = {10.1051/0004-6361/202450786},
archivePrefix = {arXiv},
       eprint = {2405.13493},
 primaryClass = {astro-ph.IM},
       adsurl = {https://ui.adsabs.harvard.edu/abs/2025A&A...697A...3E}
}

@ARTICLE{2025A&A...696A..80B,
       author = {{Bouy}, H. and {Mart{\'\i}n}, E.~L. and {Cuillandre}, J.-C. and {Barrado}, D. and {Tamura}, M. and {Bertin}, E. and {{\v{Z}}erjal}, M. and {Points}, S. and {Olivares}, J. and {Hu{\'e}lamo}, N. and {Rodrigues}, T.},
        title = "{Free-floating planetary mass objects in LDN 1495 from Euclid Early Release Observations}",
      journal = {\aap},
         year = 2025,
        month = apr,
       volume = {696},
          eid = {A80},
        pages = {A80},
          doi = {10.1051/0004-6361/202553979},
archivePrefix = {arXiv},
       eprint = {2502.16349},
 primaryClass = {astro-ph.SR},
       adsurl = {https://ui.adsabs.harvard.edu/abs/2025A&A...696A..80B}
}

@ARTICLE{2025ApJ...991...84D,
       author = {{Dominguez-Tagle}, C. and {{\v{Z}}erjal}, M. and {Sedighi}, N. and {Mas-Buitrago}, P. and {Martin}, E.~L. and {Zhang}, J.-Y. and {Vitas}, N. and {B{\'e}jar}, V.~J.~S. and {Tsilia}, S. and {Mu{\~n}oz Torres}, S. and {Lodieu}, N. and {Barrado}, D. and {Solano}, E. and {Cruz}, P. and {Tata}, R. and {Phan-Bao}, N. and {Burgasser}, A.},
        title = "{Euclid Quick Data Release (Q1){\textemdash}Spectroscopic Search, Classification, and Analysis of Ultracool Dwarfs in the Deep Fields}",
      journal = {\apj},
         year = 2025,
        month = sep,
       volume = {991},
       number = {1},
          eid = {84},
        pages = {84},
          doi = {10.3847/1538-4357/adf72d},
archivePrefix = {arXiv},
       eprint = {2503.22442},
 primaryClass = {astro-ph.SR},
       adsurl = {https://ui.adsabs.harvard.edu/abs/2025ApJ...991...84D}
}

@ARTICLE{2018AJ....156..123A,
       author = {{Astropy Collaboration} and {Price-Whelan}, A.~M. and {Sip{\H{o}}cz}, B.~M. and {G{\"u}nther}, H.~M. and {Lim}, P.~L. and {Crawford}, S.~M. and {Conseil}, S. and {Shupe}, D.~L. and {Craig}, M.~W. and {Dencheva}, N. and {Ginsburg}, A. and {VanderPlas}, J.~T. and {Bradley}, L.~D. and {P{\'e}rez-Su{\'a}rez}, D. and {de Val-Borro}, M. and {Aldcroft}, T.~L. and {Cruz}, K.~L. and {Robitaille}, T.~P. and {Tollerud}, E.~J. and {Ardelean}, C. and {Babej}, T. and {Bach}, Y.~P. and {Bachetti}, M. and {Bakanov}, A.~V. and {Bamford}, S.~P. and {Barentsen}, G. and {Barmby}, P. and {Baumbach}, A. and {Berry}, K.~L. and {Biscani}, F. and {Boquien}, M. and {Bostroem}, K.~A. and {Bouma}, L.~G. and {Brammer}, G.~B. and {Bray}, E.~M. and {Breytenbach}, H. and {Buddelmeijer}, H. and {Burke}, D.~J. and {Calderone}, G. and {Cano Rodr{\'\i}guez}, J.~L. and {Cara}, M. and {Cardoso}, J.~V.~M. and {Cheedella}, S. and {Copin}, Y. and {Corrales}, L. and {Crichton}, D. and {D'Avella}, D. and {Deil}, C. and {Depagne}, {\'E}. and {Dietrich}, J.~P. and {Donath}, A. and {Droettboom}, M. and {Earl}, N. and {Erben}, T. and {Fabbro}, S. and {Ferreira}, L.~A. and {Finethy}, T. and {Fox}, R.~T. and {Garrison}, L.~H. and {Gibbons}, S.~L.~J. and {Goldstein}, D.~A. and {Gommers}, R. and {Greco}, J.~P. and {Greenfield}, P. and {Groener}, A.~M. and {Grollier}, F. and {Hagen}, A. and {Hirst}, P. and {Homeier}, D. and {Horton}, A.~J. and {Hosseinzadeh}, G. and {Hu}, L. and {Hunkeler}, J.~S. and {Ivezi{\'c}}, {\v{Z}}. and {Jain}, A. and {Jenness}, T. and {Kanarek}, G. and {Kendrew}, S. and {Kern}, N.~S. and {Kerzendorf}, W.~E. and {Khvalko}, A. and {King}, J. and {Kirkby}, D. and {Kulkarni}, A.~M. and {Kumar}, A. and {Lee}, A. and {Lenz}, D. and {Littlefair}, S.~P. and {Ma}, Z. and {Macleod}, D.~M. and {Mastropietro}, M. and {McCully}, C. and {Montagnac}, S. and {Morris}, B.~M. and {Mueller}, M. and {Mumford}, S.~J. and {Muna}, D. and {Murphy}, N.~A. and {Nelson}, S. and {Nguyen}, G.~H. and {Ninan}, J.~P. and {N{\"o}the}, M. and {Ogaz}, S. and {Oh}, S. and {Parejko}, J.~K. and {Parley}, N. and {Pascual}, S. and {Patil}, R. and {Patil}, A.~A. and {Plunkett}, A.~L. and {Prochaska}, J.~X. and {Rastogi}, T. and {Reddy Janga}, V. and {Sabater}, J. and {Sakurikar}, P. and {Seifert}, M. and {Sherbert}, L.~E. and {Sherwood-Taylor}, H. and {Shih}, A.~Y. and {Sick}, J. and {Silbiger}, M.~T. and {Singanamalla}, S. and {Singer}, L.~P. and {Sladen}, P.~H. and {Sooley}, K.~A. and {Sornarajah}, S. and {Streicher}, O. and {Teuben}, P. and {Thomas}, S.~W. and {Tremblay}, G.~R. and {Turner}, J.~E.~H. and {Terr{\'o}n}, V. and {van Kerkwijk}, M.~H. and {de la Vega}, A. and {Watkins}, L.~L. and {Weaver}, B.~A. and {Whitmore}, J.~B. and {Woillez}, J. and {Zabalza}, V. and {Astropy Contributors}},
        title = "{The Astropy Project: Building an Open-science Project and Status of the v2.0 Core Package}",
      journal = {\aj},
         year = 2018,
        month = sep,
       volume = {156},
       number = {3},
          eid = {123},
        pages = {123},
          doi = {10.3847/1538-3881/aabc4f},
archivePrefix = {arXiv},
       eprint = {1801.02634},
 primaryClass = {astro-ph.IM},
       adsurl = {https://ui.adsabs.harvard.edu/abs/2018AJ....156..123A}
}

@ARTICLE{2020Natur.585..357H,
       author = {{Harris}, Charles R. and {Millman}, K. Jarrod and {van der Walt}, St{\'e}fan J. and {Gommers}, Ralf and {Virtanen}, Pauli and {Cournapeau}, David and {Wieser}, Eric and {Taylor}, Julian and {Berg}, Sebastian and {Smith}, Nathaniel J. and {Kern}, Robert and {Picus}, Matti and {Hoyer}, Stephan and {van Kerkwijk}, Marten H. and {Brett}, Matthew and {Haldane}, Allan and {del R{\'\i}o}, Jaime Fern{\'a}ndez and {Wiebe}, Mark and {Peterson}, Pearu and {G{\'e}rard-Marchant}, Pierre and {Sheppard}, Kevin and {Reddy}, Tyler and {Weckesser}, Warren and {Abbasi}, Hameer and {Gohlke}, Christoph and {Oliphant}, Travis E.},
        title = "{Array programming with NumPy}",
      journal = {\nat},
         year = 2020,
        month = sep,
       volume = {585},
       number = {7825},
        pages = {357-362},
          doi = {10.1038/s41586-020-2649-2},
archivePrefix = {arXiv},
       eprint = {2006.10256},
 primaryClass = {cs.MS},
       adsurl = {https://ui.adsabs.harvard.edu/abs/2020Natur.585..357H}
}

@ARTICLE{2007CSE.....9c..21P,
       author = {{Perez}, Fernando and {Granger}, Brian E.},
        title = "{IPython: A System for Interactive Scientific Computing}",
      journal = {Computing in Science and Engineering},
         year = 2007,
        month = jan,
       volume = {9},
       number = {3},
        pages = {21-29},
          doi = {10.1109/MCSE.2007.53},
       adsurl = {https://ui.adsabs.harvard.edu/abs/2007CSE.....9c..21P}
}

@ARTICLE{4160265,
  author={Hunter, John D.},
  journal={Computing in Science and Engineering}, 
  title={Matplotlib: A 2D Graphics Environment}, 
  year={2007},
  volume={9},
  number={3},
  pages={90-95},
  doi={10.1109/MCSE.2007.55}}

@INPROCEEDINGS{2005ASPC..347...29T,
       author = {{Taylor}, M.~B.},
        title = "{TOPCAT \& STIL: Starlink Table/VOTable Processing Software}",
    booktitle = {Astronomical Data Analysis Software and Systems XIV},
         year = 2005,
       editor = {{Shopbell}, P. and {Britton}, M. and {Ebert}, R.},
       series = {Astronomical Society of the Pacific Conference Series},
       volume = {347},
        month = dec,
        pages = {29},
       adsurl = {https://ui.adsabs.harvard.edu/abs/2005ASPC..347...29T}
}

@ARTICLE{2000A&AS..143...33B,
       author = {{Bonnarel}, F. and {Fernique}, P. and {Bienaym{\'e}}, O. and {Egret}, D. and {Genova}, F. and {Louys}, M. and {Ochsenbein}, F. and {Wenger}, M. and {Bartlett}, J.~G.},
        title = "{The ALADIN interactive sky atlas. A reference tool for identification of astronomical sources}",
      journal = {\aaps},
         year = 2000,
        month = apr,
       volume = {143},
        pages = {33-40},
          doi = {10.1051/aas:2000331},
       adsurl = {https://ui.adsabs.harvard.edu/abs/2000A&AS..143...33B}
}

@ARTICLE{2000A&AS..143....9W,
       author = {{Wenger}, M. and {Ochsenbein}, F. and {Egret}, D. and {Dubois}, P. and {Bonnarel}, F. and {Borde}, S. and {Genova}, F. and {Jasniewicz}, G. and {Lalo{\"e}}, S. and {Lesteven}, S. and {Monier}, R.},
        title = "{The SIMBAD astronomical database. The CDS reference database for astronomical objects}",
      journal = {\aaps},
         year = 2000,
        month = apr,
       volume = {143},
        pages = {9-22},
          doi = {10.1051/aas:2000332},
archivePrefix = {arXiv},
       eprint = {astro-ph/0002110},
 primaryClass = {astro-ph},
       adsurl = {https://ui.adsabs.harvard.edu/abs/2000A&AS..143....9W}
}

@ARTICLE{2005A&A...437..177H,
       author = {{Habart}, E. and {Abergel}, A. and {Walmsley}, C.~M. and {Teyssier}, D. and {Pety}, J.},
        title = "{Density structure of the Horsehead nebula photo-dissociation region}",
      journal = {\aap},
         year = 2005,
        month = jul,
       volume = {437},
       number = {1},
        pages = {177-188},
          doi = {10.1051/0004-6361:20041546},
archivePrefix = {arXiv},
       eprint = {astro-ph/0501536},
 primaryClass = {astro-ph},
       adsurl = {https://ui.adsabs.harvard.edu/abs/2005A&A...437..177H}
}

@ARTICLE{2025MNRAS.537..752B,
       author = {{Bate}, Matthew R.},
        title = "{Variation of the low-mass end of the stellar initial mass function with redshift and metallicity}",
      journal = {\mnras},
         year = 2025,
        month = feb,
       volume = {537},
       number = {2},
        pages = {752-778},
          doi = {10.1093/mnras/staf059},
archivePrefix = {arXiv},
       eprint = {2501.06082},
 primaryClass = {astro-ph.GA},
       adsurl = {https://ui.adsabs.harvard.edu/abs/2025MNRAS.537..752B}
}

@ARTICLE{2005ARA&A..43..195K,
       author = {{Kirkpatrick}, J. Davy},
        title = "{New Spectral Types L and T}",
      journal = {\araa},
         year = 2005,
        month = sep,
       volume = {43},
       number = {1},
        pages = {195-245},
          doi = {10.1146/annurev.astro.42.053102.134017},
       adsurl = {https://ui.adsabs.harvard.edu/abs/2005ARA&A..43..195K}
}

@ARTICLE{2026A&A...706A.210R,
       author = {{Rottensteiner}, Alena and {Petr-Gotzens}, Monika G. and {Meingast}, Stefan and {Alves}, Jo{\~a}o and {Bertin}, Emmanuel and {Bouy}, Herv{\'e} and {Piecka}, Martin and {Ratzenb{\"o}ck}, Sebastian and {Socci}, Andrea},
        title = "{Kinematics of young stellar objects in NGC 2024 based on infrared proper motions}",
      journal = {\aap},
         year = 2026,
        month = feb,
       volume = {706},
          eid = {A210},
        pages = {A210},
          doi = {10.1051/0004-6361/202557533},
archivePrefix = {arXiv},
       eprint = {2512.02124},
 primaryClass = {astro-ph.GA},
       adsurl = {https://ui.adsabs.harvard.edu/abs/2026A&A...706A.210R}
}
\bibliographystyle{aasjournalv7}



\end{document}